\documentclass[manuscript]{acmart}
\usepackage{subcaption}
\usepackage{multirow}
\usepackage{xcolor,colortbl}
\usepackage{soul}
\definecolor{light-gray}{gray}{0.9}
\definecolor{med-gray}{gray}{0.7}

\definecolor{b}{RGB}{166, 200, 255}
\definecolor{p}{RGB}{255, 175, 210}
\AtBeginDocument{%
  \providecommand\BibTeX{{%
    \normalfont B\kern-0.5em{\scshape i\kern-0.25em b}\kern-0.8em\TeX}}}

\setcopyright{acmcopyright}
\copyrightyear{2025}
\acmYear{2018}
\acmDOI{XXXXXXX.XXXXXXX}

\begin{document}

\title{Current and Future Use of Large Language Models for Knowledge Work}

\author{Michelle Brachman}
\email{michelle.brachman@ibm.com}
\orcid{0000-0001-8152-441X}
\affiliation{%
  \institution{IBM Research}
  \streetaddress{314 Main Street}
  \city{Cambridge}
  \state{MA}
  \country{USA}
  \postcode{02142}
}
\author{Amina El-Ashry}
\email{aminaelashry@ibm.com}
\orcid{0009-0000-8484-6369}
\affiliation{%
  \institution{IBM}
  \streetaddress{6710 Rockledge Dr}
  \city{Bethesda}
  \state{Maryland}
  \country{USA}
  \postcode{20817}
}
\author{Casey Dugan}
\email{cadugan@us.ibm.com}
\orcid{0000-0002-1508-2091}
\affiliation{%
  \institution{IBM Research}
  \streetaddress{314 Main Street}
  \city{Cambridge}
  \state{MA}
  \country{USA}
  \postcode{02142}
}
\author{Werner Geyer}
\email{werner.geyer@us.ibm.com}
\orcid{0000-0003-4699-5026}
\affiliation{%
  \institution{IBM Research}
  \streetaddress{314 Main Street}
  \city{Cambridge}
  \state{MA}
  \country{USA}
  \postcode{02142}
}
\renewcommand{\shortauthors}{Michelle Brachman et al.}
\renewcommand{\shorttitle}{Current and Future Use of Large Language Models for Knowledge Work}
\begin{abstract}
Large Language Models (LLMs) have introduced a paradigm shift in interaction with AI technology, enabling knowledge workers to complete tasks by specifying their desired outcome in natural language. LLMs have the potential to increase productivity and reduce tedious tasks in an unprecedented way. A systematic study of LLM adoption for work can provide insight into how LLMs can best support these workers. To explore knowledge workers' current and desired usage of LLMs, we ran a survey (n=216). Workers described tasks they already used LLMs for, like generating code or improving text, but imagined a future with LLMs integrated into their workflows and data. We ran a second survey (n=107) a year later that validated our initial findings and provides insight into up-to-date LLM use by knowledge workers. We discuss implications for adoption and design of generative AI technologies for knowledge work.
\end{abstract}





\maketitle

\section{Introduction}

The public release of Large Language Models (LLMs) has enabled widespread use of generative AI by end-users. The ability to specify a goal in natural language has the potential to transform knowledge work through automation and augmentation~\cite{IBM-Bloomberg, berg2023automation, McKinsey, parker2022automation}. In fact, one CEO stated about knowledge work in 2023: ``I could easily see 30\% of that getting replaced by AI and automation over a five-year period''~\cite{IBM-Bloomberg}. This kind of assertion motivated our work. We wanted to understand how generative AI is actually impacting knowledge work now and how knowledge workers could imagine generative AI improving their work in the future. Accounts of knowledge workers using LLMs in the workplace are limited~\cite{ritala2023transforming} and the evolution of generative AI based tools for knowledge work is just beginning (with new tools like CoPilot for MSOffice 365\footnote{https://blogs.microsoft.com/blog/2023/03/16/introducing-microsoft-365-copilot-your-copilot-for-work/}, and Zoom AI Companion\footnote{https://www.zoom.com/en/blog/zoom-ai-companion/}). To fill this gap in understanding the potential of generative AI for knowledge work, we ran a large-scale qualitative survey of knowledge workers and a follow-up study one year later \color{black} to understand their uses and needs in the space of LLM-based tools. 

Knowledge workers, who we will refer to as \textit{workers}, are those ``with high degrees of expertise, education or experience and the primary purpose of their jobs involves the creation, distribution, or application of knowledge''~\cite{davenport2005thinking}. There is a long history of research aiming to understand and support knowledge workers~\cite{kogan2006ethnographic, shen2008automatically, 10.1145/3377325.3377507, 10.1145/2166966.2166985, 10.1145/3581641.3584078}.
Despite technological advancements over the last few decades, there is still pressure for workers to improve productivity~\cite{guillou2020your}. Increased automation of tasks can improve productivity, reducing tedious and manual tasks, but also has challenges. Knowledge work is often complex and dynamic, with workflows involving multiple people and various information and data sources~\cite{kogan2006ethnographic,greene2011space}. 
Our aim was to explore both the details and context of current knowledge worker adoption of LLMs and future desired LLM-based tool use, thus providing insights into workers' needs for LLM-based tools~\cite{chilana2015user}. 

We ran a survey (n=216) to capture workers' descriptions of current and desired use of LLMs within in a large international technology company in Summer 2023 and a follow-up survey in Fall 2024 (n=107) to validate our original findings and provide insight into knowledge workers' use of LLMs one year later. Our research questions are:
\begin{itemize}
    \item RQ1: How do knowledge workers currently use LLMs for their work tasks and personal tasks, and how do they try out LLMs?
    \item RQ2: How do knowledge workers want to use LLMs in the future for their work?
    \item RQ3: How does LLM usage fit into larger and collaborative workflows?
\end{itemize}
\color{black}We used our first survey to categorize uses of LLMs through collection of a large set of descriptions of LLM usage. Our second survey enabled us to validate these categories with a new sample and establish how common the different categories of use were one year later. Our second survey establishes a baseline of use, which can be added to and replicated with additional populations in the future.\color{black}

We describe LLM usage of three groups of participants: those who had used LLMs for work (Survey1: 24.5\%, Survey2: 34.6\%), those who had used LLMs to accomplish a personal task (Survey1: 27.8\%, Survey2: 29.9\%), and those who had only experimented with LLMs (Survey1: 26.9\%, Survey2: 27.1\%)\color{black}. At a high level, we describe four categories of LLM usage for work: for creation, to find or work with information, to get advice, or for automation. Current LLM usage for work focused on creation of artifacts and ideas, finding or learning new information, and improving existing artifacts. 
In contrast, workers' personal use of LLMs involved more requests for ideas and guidance from LLMs, even on high-risk topics. 
In the future, workers hope that LLMs can also support novel insight generation, guidance and validation, automation of tasks, and further integration of LLMs into their workflows using their context and data. 
Our contributions are: 
\begin{itemize}
    \item Categorization of knowledge workers' descriptions of current LLM use for work tasks, personal tasks, and exploration, and validation and frequencies of those categories one year later\color{black}.
    \item Categorization of how workers want to use LLMs in the future to support their work  and frequencies of their desired future use one year later\color{black}. 
    \item An understanding of how current and future LLM usage fits into larger workflows.
    \item Implications for designing LLM-based systems for workers. 
\end{itemize}

\section{Related Work}
Our work contributes to research on adoption of AI technologies, intelligent agents at work, and knowledge workers.

\subsection{Adoption of LLMs}
Studying adoption has the potential to contribute to better understanding how to design technology~\cite{chilana2015user}, as well as how to mitigate users' concerns~\cite{park2021human, burgess2023healthcare}.
The recent public availability of LLM technologies, such as ChatGPT~\cite{ChatGPT}, Bard~\cite{Bard}, and GitHub CoPilot~\cite{CoPilot}, has enabled broad swaths of the public to experiment and use LLMs in a variety of contexts. Researchers have rapidly begun to explore early perceptions of LLMs, primarily on the usage of LLMs in domains like education~\cite {bonsu2023consumers, li2023chatgpt, dahlkemper2023physics, shoufan2023exploring, chan2023students}, healthcare~\cite{praveen2023understanding, zheng2023innovating}, and tourism~\cite{carvalho2023chatGPT, demir2023professionals}. We focus our review of related work on: 1) populations of adopters and 2) ways people use LLMs currently and in the future.

\subsubsection{Populations of adopters}
Studies of populations' LLM adoption provide early insights into the demographic and experiential qualities of LLM users and provide insight into workers' use of LLMs. 
A study of visitors to ChatGPT's website showed that ChatGPT activity was associated with lower age, more education, and living in a rural area~\cite{kacperski2023users}, though another survey showed similar ChatGPT usage across generations~\cite{ChatGPTCatchingOn}. 
Several surveys have recorded proportions of the public using LLMs at work, with one showing 25\%~\cite{ChatGPTCatchingOn} and another showing 38\% of people using LLMs at work~\cite{skjuve2023user}. Twitter data also gives some insight into the kinds of workers, such as software developers and data scientists, that are discussing LLM usage on social media~\cite{haque2022exploring}. Yet, several studies have shown that job role has a low or weak impact on use of LLMs~\cite{agossah2023llmbased, makkonen2023effects}. Instead of focusing on the qualities of adoption groups, we look at use of LLMs by workers of three adoption groups to shed light on LLM usage for work: those who use LLMs for work, personal use, and for just trying out LLMs. 

\subsubsection{Types of usages}
Research has identified the way people use LLMs and generative AI in a variety of contexts, such as education~\cite{yan2023practical}, healthcare~\cite{zheng2023innovating}, writing~\cite{gmeiner2023dimensions}, and personal usage~\cite{taecharungroj2023can, skjuve4376834people}.  
Closest to our work, an interview study of 22 information workers in the spring of 2023 found that knowledge workers were using ChatGPT for: answering questions, serving as a search engine, generating content, improving content, generating and improving code, supporting learning, and handling emails and reminders~\cite{ritala2023transforming}. A study of an LLM programming assistant showed that users were using the tool for tasks like explaining code and recalling syntax, while they wanted to use the tool for tasks like refactoring and getting suggestions for code improvements ~\cite{ross2023programmer}.
We support and expand upon knowledge worker LLM usage, considering how these usages are part of larger workflows, exploring how different adoption groups use LLMs differently, and exploring how knowledge workers want to use LLMs to augment and automate their work in the future. Capturing these future desires and expectations also allows tracking against them - both on the AI side (whether and when the LLMs are able to do as knowledge workers envision) and on the human side, whether humans fail to follow through with their anticipated adoption for any number of reasons (changing desires, public sentiment, legislation, etc).

\subsection{Conversational Agents, Chatbots, and Intelligent Agents at Work} 
Though LLMs have many applications and potential usages, ChatGPT~\cite{ChatGPT} and the nature of text generation makes conversational agents a natural application of LLMs. Conversational agents, chatbots, and intelligent agents enable users to interact through natural language with a system to accomplish a variety of goals, such as information search~\cite{vtyurina2017exploring}, data analysis~\cite{fast2018iris}, and coding~\cite{ross2023programmer}. These systems have applications in a variety of domains, such as in healthcare~\cite{laranjo2018conversational}, education~\cite{khosrawi2022conversational}, as well as customer-facing business contexts~\cite{xu2017new}. Our work is most related to literature on existing use and perceptions of conversational agents or virtual enterprise assistants for knowledge workers~\cite{gkinko2020creation}. 
Research has primarily focused on two dimensions of chatbots in the workplace: intention to use and perceptions, like emotion. The perceptions and use of chatbots and intelligent assistants in the workplace is highly personal~\cite{drebert2023influence} and contextual~\cite{he2023understanding}.
For example, use of a chatbot for IT workers was impacted by users' understanding of the tool and use of the chatbot in rational vs. emotional ways~\cite{gkinko2023appropriation}. 
Our study focuses on the scenarios where workers currently use LLMs (often in the form of conversational agents like ChatGPT) and scenarios where they would like LLM-based support in the future. Our results may apply to the design of conversational agents for knowledge workers, but are not specific to conversational systems, suggesting spaces where knowledge workers can imagine LLM-based support.


\subsection{Enterprise Knowledge and Information Workers} 
Our work focuses specifically on knowledge workers' current and desired use of LLMs. 
Researchers have found a variety of challenges in knowledge work. 
Work tasks are often part of larger workflows, which are often implicit, requiring mining to capture them~\cite{van2003workflow, di2013mining, van2003weakly}.
Within a workflow, work is often interrupted, requiring people to leave and return to contexts~\cite{mark2005no, bannon1983evaluation}. These contexts require many digital artifacts and resources, which workers utilize for individual tasks and across workflows~\cite{bannon1983evaluation, hu2022scrapbook}. Further, knowledge work is often collaborative~\cite{houben2013activity, iivari1999knowledge, mundbrod2013towards,kogan2006ethnographic} and requires communication between individuals, such as through email, which can add even more strain to an already fragmented work context~\cite{dabbish2006email, whittaker1996email}. These challenging aspects of modern knowledge and information work can lead to stress and well-being concerns~\cite{mark2008cost}. Researchers have aimed to support workers in finding and handling resources~\cite{shen2008automatically}, coordination of work~\cite{houben2013activity}, email management~\cite{10.1145/1943403.1943434}, skill acquisition~\cite{10.1145/2557500.2557539}, well-being~\cite{10.1145/3411764.3445388}, and productivity~\cite{10.1145/3544548.3581326}. However, challenges and annoyances remain in knowledge work, as well as a need to better understand how to support productivity~\cite{palvalin2019matters}. Inherent in the ways knowledge workers use and would like to use LLMs are spaces where they would like support or to improve productivity. Our work contributes to the understanding of knowledge workers by documenting the ways workers are currently attempting to use LLMs for work and how they would like to use LLMs in the future.

\section{Methods}
\subsection{Survey}

\subsubsection{Design Process}
We chose to use a survey for this research because we wanted to capture a large number of responses across a broad population of workers.  For Survey1, \color{black} since the availability of LLMs for the general public was still relatively new, we also wanted to explore how workers described their current and future uses of LLMs in their own words, rather than asking them to categorize their use. Thus Survey1 \color{black} contains primarily open-ended questions and some multiple choice and Likert-scale questions. We focused our survey on LLMs with text inputs and outputs (such as ChatGPT, Bard, etc), rather than broader categories of foundational or generative AI in order to limit the scope.  For Survey2, we stayed as close as possible to the survey design in Survey1, but wanted to validate our themes from Survey1 \color{black} and establish a baseline of LLM usage in knowledge work\color{black}. Thus, Survey2 asked participants to select their use of LLMs using the themes generated from the analysis of Survey1. \color{black}Like Survey1\color{black}, Survey2 also asked participants to \color{black}describe \color{black} current and future \color{black}desired \color{black}use of LLMs\color{black}. To capture a diverse sample of LLM usage descriptions, Survey2 requested the description for \color{black} a randomly selected use type.\color{black}

We were inspired by the activity checklist~\cite{kaptelinin1999methods} to design questions that addressed the larger context of LLM usage. 
The activity checklist has four main concerns about an activity: means and ends, environment, learning, and development. Means and ends is about understanding whether a technology supports users in accomplishing their goals, such as our questions about knowledge workers' background and their goals in using LLMs. Environment is about considering how a technology integrates with other features of the larger context like requirements, tools and resources. Our survey asks about the larger context and workflow around LLM usage. Learning is about the process of understanding how to use a technology and problem solving, while development is about how activities and needs change. To address learning, our survey asks about users' frustrations in using LLMs. To address development, we asked how workers would like to use LLMs in the future, and how they previously accomplished tasks that they have begun to use LLMs for. The activity checklist has a set of questions that we used for inspiration, modifying them to fit our context. 

Two authors iterated on the survey questions, including discussion with two other authors and five pilot participants before finalizing the questions. Pilot tests showed that the survey took about 15-20 minutes to complete. We iterated on the length of the survey, as well as details of the questions, such as adding the option to categorize adoption of LLMs as ``just trying it,'' rather than just use at work or for personal use because participants ended up selecting that they didn't use LLMs at all if they did not feel that their exploration of LLMs was significant enough.

\subsubsection{Survey Questions}
\begin{figure}[bt]
\caption{High-level survey flow, with topics discussed based on question of how they used LLMs.}
\label{fig:survey}
\includegraphics[width=\linewidth]{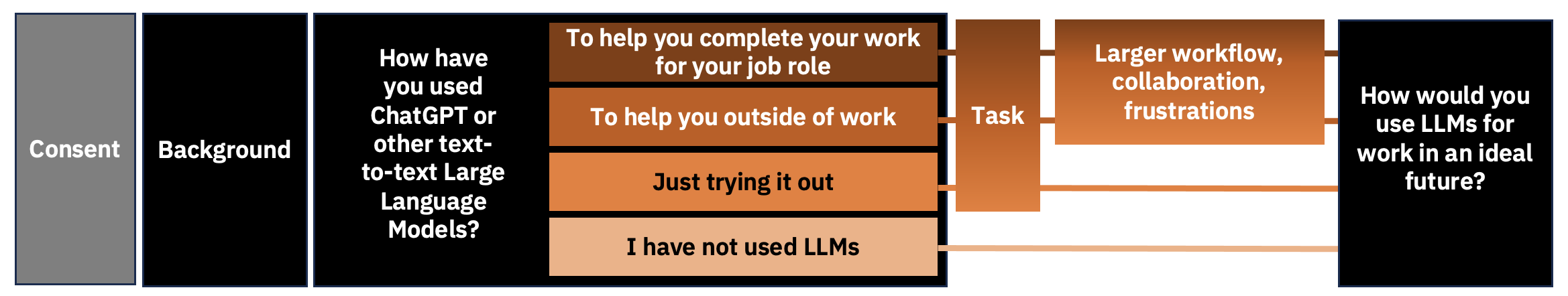}
\end{figure}

After consent, both surveys began with a page collecting background data about the participants, including their job role, work location, experience with programming and AI, and trust in AI~\cite{jessup2019measurement}. We asked participants about their general AI experiences to capture whether they may have used other related systems, like non-LLM chatbots or personal assistants, which may impact their trust in AI and/or ease of use with AI. 
It then asked participants questions about their adoption of LLMs (use for work, use for personal, only trying it out, or no use, as shown in Figure~\ref{fig:survey}).
Participants could select all options that applied, but if they selected that they used it for work, they went down the work-use path, regardless of whatever else they selected. If they did not select work, but did report that they used an LLM for a non-work task, the survey directed them down the non-work path. If they did not select work or non-work LLM use, but did report that they had tried out an LLM, they were sent down the tried-it path. Finally, participants who had not used LLMs for other purposes followed the no-use path.
Participants who had used LLMs for work or personal tasks received a similar set of questions, probing at details of the tasks, while those who had only tried out LLMs or had not used them at all had similar sets of questions. 
Finally, the survey asked all participants about their idealized future uses of an LLM at work. 
The main difference in questions between Survey1 and Survey2 is that the questions about use of LLMs were \color{black} entirely \color{black}open-ended in Survey1. In Survey2, participants selected their use of LLMs based on the types of LLM usages found in Survey1 and then provided examples of those selections. There were some other minor changes in the demographic questions due to company policies (see Appendix Tables ~\ref{tab-demo} and ~\ref{tab:demo2}) and some questions were removed for brevity in Survey2 \color{black}that weren't the focus of this work, such as around personal use of LLMs\color{black}. One question was added about challenges in using LLMs for larger workflows or collaboration \color{black} to further probe the potential future needs of knowledge workers using LLM-based systems. \color{black}
The survey questions for both surveys \color{black} are available in the supplementary material.



\subsection{Participants}

We report participant recruiting and demographics for Survey1 (S1) and then Survey2 (S2).
\subsubsection{Survey1\color{black}}
\color{black}
We recruited participants broadly across a large international technology company through posts in over 80 internal Slack channels from June 23, 2023 to August 1, 2023. The company had restrictions on usage of some LLM technologies for work uses due to privacy concerns. However, the company does create LLMs and LLM technologies, which were available to employees during this time. 
We aimed to recruit across a wide range of technical experience, AI experience, job roles, and geographic regions to better support generalizability of results. Further, the AI focus of the company and high AI/technology exposure of the knowledge workers likely makes them more able to appropriately speculate about how they might want to use LLMs in the future. 

In total, 380 participants started our survey, 45 dropped off during the background questions, and 119 participants dropped off during the questions that asked for more detail on use of LLMs. We analyzed the data from 216 participants who completed the survey. We stopped recruitment based on slowing of new participation. \color{black}
Our participants were primarily from North America (63\%), Europe (22\%), or Asia (10\%). While many participants had at least partially technical roles (45\%), more than half were non-technical, with roles like design/UX (19\%), sales (16\%), marketing (11\%), and \color{black}customer service (7\%). \color{black}

Participants had a range of technical backgrounds and trust in AI. They had a wide range of programming experience (Mean= 5.1, SD= 3), rated on a 10-point scale~\cite{siegmund2014measuring}. Their trust in AI was also varied (Mean= 2.8, SD= 1), rated on a five-point scale~\cite{jessup2019measurement}. Participants selected options for their experience with AI, such as ``Tried consumer AI tools,'' and ``Significant AI work experience.'' \color{black} 
Overall, only 12\% of participants had significant AI experience, meaning that they work on training or development of LLMs. See Table~\ref{tab-demo} in the Appendix for further details on percentages of participants per continent, AI experience, and job responsibility. 
Our participants were roughly evenly split between having used LLMs for work (24.5\%), for personal use but not work (27.8\%), having tried out LLMs (26.9\%), and not having used LLMs at all (20.8\%).
Table~\ref{tab:use} shows the breakdown of overlapping use between these groups. For example, we group all participants who had used LLMs for work together (\textit{work-use}), as they went through the survey the same way, even though some had also used LLMs for personal use and others had not. \color{black}
To give a better understanding of the makeup of these groups, we report the average programming experience and percentage of participants in each adoption group with significant AI work experience (see Table~\ref{tab:use}, last two columns). 

\begin{table}[tb]
\caption{Adoption groups, along with average programming experience and percentage of each group with significant AI work experience for Survey1 (S1, n=216) and Survey2 (S2, n=107)\color{black}. }~\label{tab:use}
\small
\begin{tabular}{p{.5cm}p{.75cm}p{.5cm}p{.75cm}p{1.5cm}p{2.5cm}p{1.5cm}|p{2cm}p{1.5cm}}
\toprule
Used for work & Used outside of work & Tried it & Haven't used & Count  (S1 total: 216, S2 total: 107) \color{black}& Percentage &Group Name & Programming Exp. (1-10) & \% Significant AI work experience \\ \hline
\checkmark & \checkmark & \checkmark & - 
        & S1:\color{black}25 S2:10 &  \multirow{4}{2.5cm}{S1:\color{black}24.5\%  S2: 34.6\%\color{black}} 
        & \multirow{4}{*}{\textit{work-use}} & \multirow{4}{2cm}{ S1:\color{black}M=6.1,SD =3.2 S2:M=4.3,SD=2.8}& \multirow{4}{1.5cm}{ S1:\color{black}26\% S2:14\%}\\
\checkmark & \checkmark & - & -&  S1:\color{black}20 S2:19& & \\
\checkmark &-  & \checkmark & -& S1:\color{black}4 S2:4& & \\ 
\checkmark &-  & - &- & S1:\color{black}4 S2:4& & \\\hline

-& \checkmark & \checkmark & -
        & S1:\color{black}41 S2:10 & \multirow{2}{2.5cm}{S1:\color{black}27.8\% S2:29.9\%} 
        & \color{black}\multirow{2}{*}{\textit{personal-use}} & \multirow{2}{2cm}{S1:\color{black}M=4.6,SD=3.3 S2:M=4.8,SD=3.1}& \multirow{2}{1.5cm}{S1:\color{black}12\% \hspace{1mm}S2:9\%}\\
-& \checkmark &- & -&   S1:\color{black}19 S2:22 & & \\\hline

- & - & \checkmark &-
        & S1:\color{black}58 S2:29 & S1:\color{black}26.9\% S2:27.1\%
        & \color{black} \textit{tried-it} &  S1:\color{black}M=5.2,SD=3 S2: M=5.6,SD=3.7 &  S1:\color{black}3\% \hspace{2mm}S2:0\%  \color{black}\\ \hline

\color{black}-& - & - & \checkmark 
        &  S1:\color{black}45 S2:9 &  S1:\color{black}20.8\% S2:8.4\% 
        & \color{black}\textit{no-use} &  S1:\color{black}M=4.5,SD=3.2 S2:M=5.2,SD=3.9& S1:\color{black}7\% \hspace{2mm}S2:0\%\color{black}\\
\bottomrule
\color{black}
\end{tabular}
\color{black}
\end{table}

\subsubsection{Survey2}
We recruited participants from September 11, 2024 to October 17, 2024 using the same methods as in Survey1. We had 107 participants complete the survey. Due to internal policies, we had to collect demographic information slightly differently. However, our demographics were largely similar to Survey1. We had 52\% of participants from the Americas, 39\% from Europe, the Middle East and Africa, and 8\% from the Asia Pacific region. Participants had an average programming experience of 4.8 (SD = 3.2) on a 10-point scale. On average, their trust in AI was  2.9 (SD = 1) on a 5-point scale. Most participants had some, but not extensive AI experience, with only 4\% of participants having significant work experience and only 8\% of participants not knowing much about it. 
See Table~\ref{tab:demo2} in the Appendix for further details on percentages of participants per geographic region, AI experience, and job responsibility for Survey2. 
\color{black}


\subsection{Data and Analysis}

We performed a qualitative analysis of our open-ended responses from the 216 participants  from Survey1 \color{black} to categorize the ways participants use and want to use LLMs. For the open-ended questions, we performed an inductive reflexive thematic analysis~\cite{braun2006using, braun2021one}. We chose an inductive approach, as our study was exploratory and need-finding in nature. 
We followed the six phases of thematic analysis: 1) familiarization with the data, 2) generate initial codes, 3) search for themes, 4) review themes, 5) define and name the themes, and 6) produce the report~\cite{braun2006using}.
Two authors familiarized themselves with the data by reading through it multiple times and taking notes. They then generated initial codes by passing through the data. They sorted the codes into themes and collected the data for those themes. They then reviewed the themes together, checking that the data fit into the themes and revising themes when needed. The authors then defined and named themes by looking at the data in the  theme, where a theme: ``captures something important about the data in relation to the research question and represents some level of \textit{patterned} response or meaning'' ~\cite{braun2006using}. 
Due to the inductive and reflexive nature of our work and to remain faithful to the reflexive thematic analysis method, we did not aim to establish inter-rater reliability but instead acknowledge the impact of the authors on the analysis~\cite{braun2006using, gauthier2022will, miller2022barriers,renney2022studying}. Table~\ref{tab:task-types} shows our themes and sub-themes along with descriptions. 

For Survey2, we provide the frequencies of selected LLM usage types and associated quotes, which are linked directly to the themes by the Survey2 design (participants select their usages and then are asked to provide an example of a specific LLM usage type that they selected). 
\color{black}



\section{Results}
We answer our three research questions: 1) How do knowledge workers currently use LLMs for their work tasks and personal tasks, and how do they try out LLMs, 2) How do knowledge workers want to use LLMs in the future for their work, and 3) How does LLM usage fit into larger and collaborative workflows?
We describe participants by their highest level of adoption (in order from work to no-use, where we believe work use is more adoption than personal use), as many participants selected multiple ways they have used LLMs.

\subsection{RQ1: How do knowledge workers currently use LLMs for their work tasks and personal tasks, and how do they try out LLMs?}
We wanted to understand the kinds of tasks workers described currently using LLMs for. All participants who reported using LLMs answered this question, though in Survey1 the question was open-ended (171/216 Survey1 participants: 53 work-use, 60 personal-use, 58 tried-it) and in Survey2, participants selected the task types we discovered in Survey1 (98/107 Survey2 participants: 37 work-use, 32 personal-use, and 29 tried-it). Figure~\ref{fig:survey1-counts} shows the frequency of task types described in Survey1 and Figure~\ref{fig:survey2-counts} shows the frequency of task types selected in Survey2. \color{black}
We describe current LLM usages across four \color{black} categories: creation, information, advice, and automation.
In Survey2, participants selected from the categories we found from Survey1, but could also choose other and write in a response. However, only three participants selected ``other'' (testing ideas, social selling, and consolidating information), indicating that our task categories cover the majority of existing tasks. 
\color{black}

\begin{table}[]
\caption{Themes and sub-themes for LLM usage }~\label{tab:task-types}
\small
\begin{tabular}{llp{8cm}l}
\toprule
Theme & Sub-Theme & Description & Current, Future, or Both \\
\hline
\multirow{2}{*}{Creation} & Artifact & Generate a new artifact to be used directly or with some modification (such as an email, code, a blog post, marketing content) & Both \\
& Idea & Generate an idea, to be used indirectly & Both \\ \hline
\multirow{2}{*}{Information} & Search & Seek a fact or piece of information & Both \\
& Learn & Learn about a new topic more broadly (like Kubernetes)& Both \\
& Summarize & Generate a shorter version of a piece of content that describes the important elements. & Both \\
& Analyze & Discover new information about a set of information or data. & Both \\ \hline
\multirow{3}{*}{Advice} & Improve & Generate a better version (like improved grammar or improved code) & Both \\
& Guidance & Get guidance about how to make a decision.& Both \\
& Validation & Check whether an artifact satisfies a set of rules or constraints. & Both \\ \hline

Automation& Automation & Complete a task in a piece of software with less or no human effort. & Future \\

\bottomrule
\end{tabular}
\end{table}

\subsubsection{Creation}
\begin{figure}
\captionsetup{labelfont={color=black}}
    \centering
    \textcolor{black}{\fboxrule=0pt\fbox{\includegraphics[width=\linewidth]{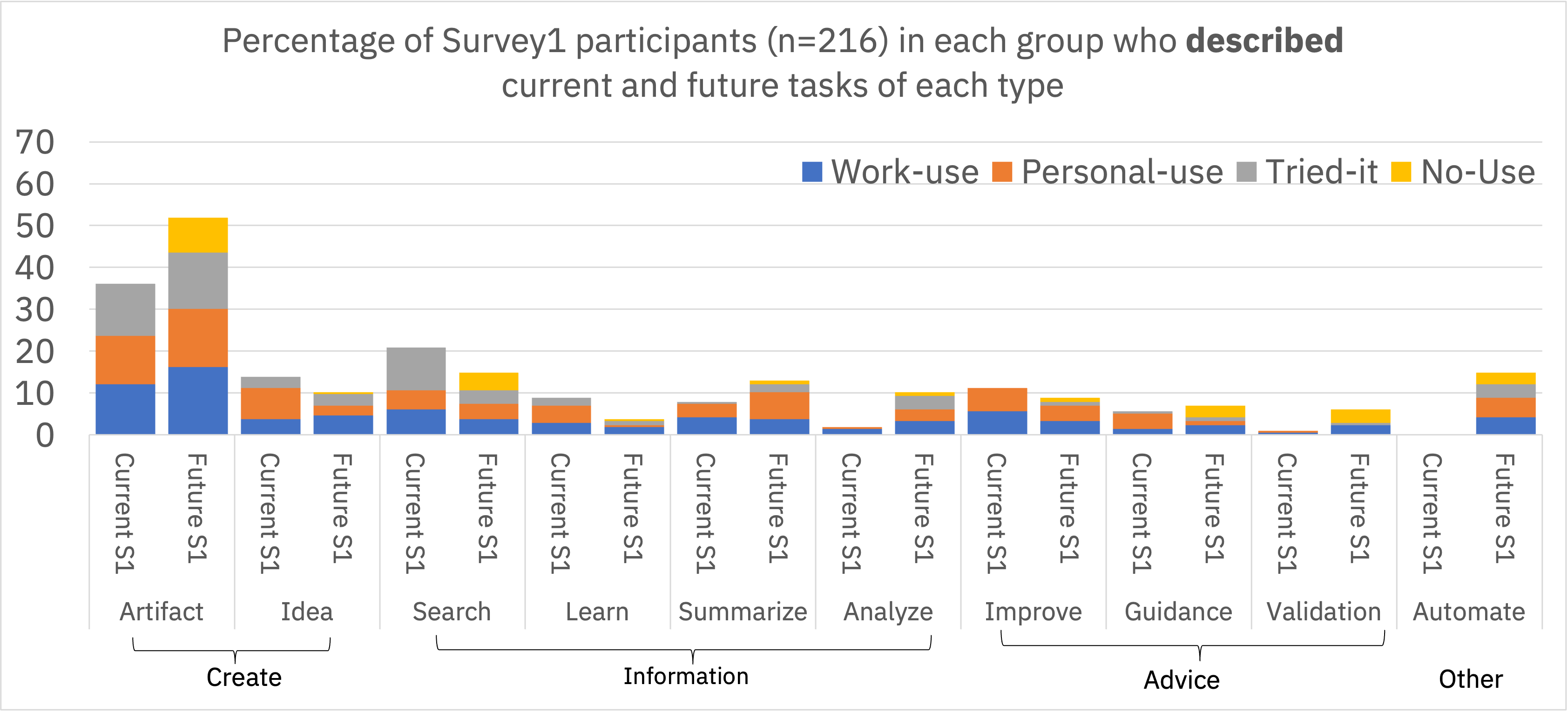}}}
    \caption{Frequency of Survey 1 task types, based on inductive analysis of participants' open-ended descriptions of the tasks they currently use LLMs for and tasks they would like to use LLMs for in the future. Note: participants were \textbf{not} asked to describe all tasks, so these frequencies of description cannot be directly compared to Survey2.}
    \label{fig:survey1-counts}
\end{figure}

\begin{figure}
\captionsetup{labelfont={color=black}}
    \centering
    \textcolor{black}{\fboxrule=0pt\fbox{\includegraphics[width=\linewidth]{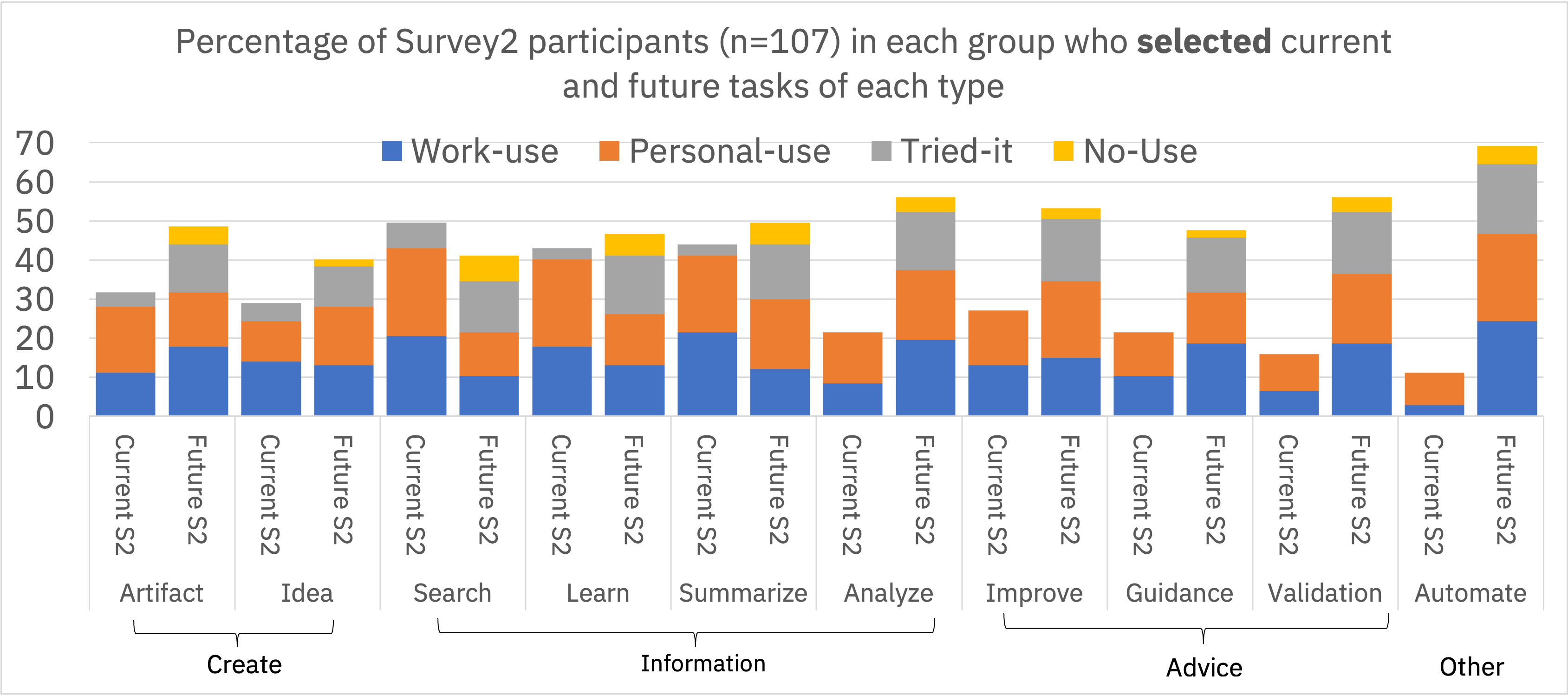}}}
    \caption{ Frequency of Survey2 task types, based on multiple choice (select all) questions asking which types of tasks participants currently use LLMs for and which tasks they would like to use LLMs for in the future.}
    \label{fig:survey2-counts}
\end{figure}

There were two main ways users wanted LLMs to help them in creation: text they could either use or modify for their goal (creation-artifact) or ideas that they could use to further their thinking on a topic (creation-idea).  In Survey1, artifact creation was the most frequently described way participants were using LLMs across all groups (work-use, personal-use, and tried-it, see Figure~\ref{fig:survey1-counts}). In Survey2, participants continued to select artifact creation and idea creation as current uses of LLMs and described similar scenarios as Survey1, but creation was not the most common use case selected (see Figure~\ref{fig:survey2-counts}). 
\color{black}
As expected, participants frequently described generating technical artifacts, like code or commands. P135-s1work \color{black} said that they used an LLM \emph{``to generate unit test with lots of mock data for my code.''} Current creation requests are often discrete, like SQL commands (P231-s1work\color{black}) or an Excel script (P223-s1work\color{black}). 
For non-technical work, participants used LLMs for \emph{```starter' draft responses [for a chatbot] that I mod in brand voice and edit for accuracy''} (P198-s1work\color{black}) and \emph{``to get me started on writing a job req [job requisition], removing what would be akin to writers block''} (P255-s1work\color{black}). 
Importantly, participants discussed these generations as starting points or drafts, such as P274-s1work \color{black} who wrote: \emph{``The code snippet didn't work as-is, but it was helpful as a starting point for my code.''}
Workers also sought novel ideas or brainstorming support from LLMs. P69-s1work \color{black} said: \emph{``We asked ChatGPT to brainstorm a list of design/research skills that would be helpful to help grow our team.''} Other participants sought persona ideas (P54-s1work\color{black}) and title and keyword ideas (P172-s1work\color{black}).
However, participants faced challenges, like in using LLMs for creative text generation: \emph{``My main challenge is that, in doing creative work, the platforms I use try to make everything so neat and tidy, like a boring human trying to sound "correct" but not really that good or interesting''} (P143-s1work\color{black}).

Those who used LLMs for personal use or had only tried out LLMs completed similar tasks as work use, but focused on using LLMs to generate ideas and creative text. Personal-use participants sought ideas throughout their daily life, such as \emph{``travel itineraries''} (P60-s1personal\color{black}), \emph{``cooking recipe ideas''} (P61-s1personal\color{black}), \emph{``workout schedules''} (P220-s1personal\color{black}), and \emph{``a list of options for my kid to dress as for cultural day at school''} (P253-s1personal\color{black}). Those who only tried out LLMs requested creative generations, such as \emph{``Writing children's stories for amusement purposes''} (P141-s1tryit) or \emph{``Generate Father of the bride speech; Write a Eulogy''} (P271-s1tryit). While there may be some overlap in needs between personal and work creation uses, such as needing creative capabilities, work use tasks often require knowledge that is specific and detailed, while non-work tasks use more common knowledge, such as cooking or travel. 

\subsubsection{Information}
Participants often described using LLMs to obtain or understand information by: finding particular pieces of information (search), gaining new information about a topic (learn), obtaining a concise description of longer content (summarize), or gathering new insights about information or data (analyze). 
 In Survey1, search was the most commonly described current information task, especially among tried-it participants (see Figure~\ref{fig:survey1-counts}). In Survey2, information tasks were the most frequently selected current uses of LLMs, with search, learn, and summarize all very common (see Figure~\ref{fig:survey2-counts}). One space of progress in the last year has been context sizes, which continue to increase, enabling more text to be input in a prompt. This may be explain the high frequency of summarization uses in Survey2. Another space of improvement for information tasks is systems that use tools like web search to find and reference information sources (such as Perplexity~\footnote{https://www.perplexity.ai}). These systems may make information search and learning more grounded in content that can be verified, in contrast to LLM-based tools that do not interact with or reference external information. \color{black}

Despite the known issue of LLMs providing incorrect responses that sound correct~\cite{shen2023chatgpt}, participants still used LLMs in place of search even for specialized work topics. For example, P257-s1work \color{black} wrote: \emph{``Getting answers on security topics instead of googling them.''} 
Participants also used LLMs more broadly to find information like \emph{``research on a topic''} (P186-s1work\color{black}). One participant described their use of an LLM for learning in detail: \emph{``Once I heard we are about to start working with Kubernetes, I immediately started a chat with ChatGPT about everything pertaining to it. It summarizes really well, and it matches itself to me in terms of complexity and professionality. That is, it understands that I am not a novice in programming, and that I have some experience with cloud, and it explains to me in the right level, neither too condescending, nor too over my head''} (P174-s1work\color{black}). While this is useful in some cases, the same participant noted that they found a lack of knowledge about REST APIs, with the LLM even providing examples from a wrong API. Finally, work-use participants used LLMs typically to summarize lengthy public content, like P258-s1work \color{black} wrote: \emph{``Summarize/text extract from long-winded public website pages.''} Analysis was rare in current LLM uses, but will be discussed further in participants' desires for future uses of LLMs. Both summarization and analysis have limitations, such as token limits on large documents and data that can make it challenging for an end-user to use LLMs for these use cases (P322-s1work\color{black}). 

Outside of work, tried-it participants described asking questions as a way to evaluate the quality of LLMs. For example, P108-s1tryit \color{black} described how they experimented with LLMs as: \emph{``basics, like answering questions.''} 
In these cases, participants already knew the answers and were testing whether the LLMs could provide the correct answers and the types of knowledge they had. Thus, workers are trying to evaluate the capabilities of LLMs, though they may not be doing this in systematic or thorough ways. 

\subsubsection{Advice}~\label{secction:curr-advice}
Workers are using LLMs to obtain three kinds of advice: 1) improve, in which the LLM takes a user's artifact and makes it better, 2) guide, in which the LLM makes a recommendation about how to proceed, and 3) validate, in which the LLM checks whether an artifact fulfills a set of requirements. 
 In Survey1, improve was the most commonly described current advice task, with few describing guidance or validation tasks (see Figure~\ref{fig:survey1-counts}). In Survey2, advice tasks remained less frequently selected than create and information tasks, with improve still the most common (see Figure~\ref{fig:survey2-counts}). 
\color{black}
Current LLM usage focuses on improvement of artifacts, often generally improving writing or for email. P33-s1work \color{black} wrote that they used an LLM to: \emph{``help me improve my writing (e.g. make a sentence more concise or coherent, come up with a more suitable word to describe something).''} However, even in seemingly simple cases like grammar fixes, participants did not always get back what they expected: \emph{``get frustrated if I ask only fix grammar and got back the paragraph with changed meaning''} (P123-s1work\color{black}). Several work-use participants also used LLMs for validation or guidance with respect to bugs or errors in code or software. 
In Survey2, participants had begun using LLMs for more complex advice, such as code validation: \emph{``I employ [anonymized tool] to analyze code to verify that the function does what it is supposed to, and to generate some basic unit tests. I still need to do most of the legwork myself, but it does catch some things quicker when it works''} (P59-s2work).
\color{black}

In contrast, outside of work, participants requested specialized guidance from LLMs on topics that typically require expertise and personalized context. For example, participants had used LLMs for \emph{``diagnosing a respiratory infection and researching treatment''} (P37-s1personal\color{black}), parenting advice (P47-s1personal\color{black}), and advice for purchasing a house (P87-s1personal\color{black}). Considering known risks with outputs as well as privacy concerns, we were surprised that workers were using LLMs for such personal and specialized areas of advice.

\subsubsection{Automation}
 In Survey1, no users described using LLMs for automation tasks, but in Survey2, several participants selected that they currently use LLMs for automation (though if participants selected multiple current uses, we randomly asked for one example, so we are lacking descriptions of these automation cases). This suggests that knowledge workers are beginning to have access and ability to fulfill their desired uses of LLMs from Survey1. 
\color{black}

\subsection{RQ2: How do knowledge workers want to use LLMs in the future for their work?}

All participants responded to this question, but only 181/216 Survey1 participants described a potential future use of LLMs at work (51/53 work-use, 50/60 personal-use, 46/58 tried-it and 34/45 no-use). The other participants wrote that they would not want to use LLMs in the future for work or could not think of a use. All but 4 Survey2 participants (103/107) selected ways they would want to use LLMs in an ideal future at work (36/37 work-use, 29/32 personal-use, 29/29 tried-it, 9/9 no-use). Figure~\ref{fig:survey1-counts} shows the frequency of future task types described in Survey1 and Figure~\ref{fig:survey2-counts} shows the frequency of future task types selected in Survey2. \color{black}
We describe desired LLM usages across four categories: creation, information, advice, and automation.  In terms of frequency, in Survey1, participants described desired future LLM tasks that primarily involved creation of artifacts. However, in Survey2, the selection of desired future tasks was much more evenly spread, with automation as the highest future task, followed by validation and analyze. \color{black} To avoid repetition, we focus on future desired uses that participants did not describe as part of current work practices (above in RQ1). 

\subsubsection{Creation}
Imagined future uses of LLMs are complex, integrated into work, and include specialized skills beyond code. 
Participants' future visions of LLM usage also included more support for email and message creation. Participants described wanting to generate documentation and tests for code (P309-s1personal\color{black}) and spreadsheet building (P38-s1nouse\color{black}), rather than the code snippets users currently generate. These both may need LLM integration directly into software products and access to significant amounts of data and information. 
Participants also want even further help with creating artifacts that are central to their job roles, like papers (P34-s1work\color{black}), \emph{``epics and user stories''} (P282-s1work\color{black}), \emph{``sales or training presentations''} (P263--s1tryit\color{black}), reports (P260-s1work\color{black}), and performance reviews (P340-s1tryit\color{black}). Yet, participants still talk about using LLMs for \emph{``a first draft''} (P353-s1nouse\color{black}) and for writing the \emph{``fluffy part of documents''} (P243-s1nouse\color{black}).

Participants in Survey2 had some similar desires for future tasks, like writing emails (P27-s2work), code (P103-s2work), and standards documents (P77-s2work).
\color{black}

\subsubsection{Information}
In the future, participants wanted to be able to perform information tasks using their own data, especially searching, summarizing, and analyzing. 
In an ideal future, participants want to use LLMs to search within their own data, such as \emph{``I would love to train a bot with all of the guild's content so I can offer a bot to answer FAQs on Slack.''} (P140-s1personal\color{black}), or \emph{``I would like to be able to upload all of our source code, and the companies policies and best practices into a LLM. Then I could ask it questions when other developers are gone for the day or have left the company.''} (P39-s1work\color{black}). These scenarios involve more than merely a basic input and output interaction with an LLM due to the large amounts of data. 
 In Survey2, participants thought about more complex search capabilities, like \emph{``as the LLM is able to deduce related searches it can at the same time broaden the search field and narrow it down to the intended meaning''} (P94-s2work), and \emph{``searching forums for a solution tailored precisely for my query''} (P39-s2work).
\color{black}

Participants also described focused summarization needs, such as support with meeting minutes (P329-s1personal\color{black} , P87-s2work\color{black}) and emails (P180-s1personal\color{black}).
 P100-s2work also described wanting summaries of \emph{``more complex items including architecture, proposals,''} indicating that the existing summarization capabilities may not work well for these more complex documents.
\color{black}
They also described specialized and personalized LLM analysis usages, like \emph{``data from Jira to create an on-the-glass view based on multiple projects across multiple dashboards to show me one view with status updates, milestone dates''} (P98-s1personal\color{black}). The participant in this case is likely describing a system that includes more components than just an LLM, but is imagining a future system that can use an LLM to effectively analyze information about projects (imagine it analyzing multiple ``boards'' that include project management information) and potentially using other components to merge the information and provide a summary view.
Participants seemed more interested in using LLMs for summarizing and analysis in the future and described using LLMs integrated into their data and workflows. 

\subsubsection{Advice}
For the future, users described scenarios including validation and guidance based on their data. P135-s1work \color{black} hoped that LLMs can validate their code: \emph{``Do code reviews that analyze the pull request description to check if it's properly worded, and to see if the description matches the actual changes in the files.''}
 In Survey2, P59 talked about a similar current use, as described in Section~\ref{secction:curr-advice}.
\color{black}
P33-s1work \color{black} wants suggestions on demand: \emph{``I work with a lot of visual artifacts (e.g. prototypes, presentation slides, workshop templates). It might be helpful if an LLM could be embedded into those tools and suggest ideas or improvements when I ask for them.''} Participants also wanted to be able to request feedback on other specialized work, like user research tasks (P110-s1personal\color{black}). 

In Survey2, participants also described complex advice needs that integrate with their data, such as analyzing logs (P41-s2work), getting different possible solutions for an issue (P1-s2work), or getting a vision for how to improve a proposal (P19-s2work). 
\color{black}

\subsubsection{Automation}
Participants described future scenarios where LLMs would perform tasks for them. 
Participants often described automation as a desired use in Survey1 and in Survey2, nearly 70\% of participants selected automation as a desired future use of LLMs. \color{black}
Automation often requires access to APIs or control over user interfaces to perform. For example, participants described wanting LLMs to \emph{``handle my calendar''} (P186-s1work\color{black}), which could involve several tasks like scheduling meetings, and responding to meeting requests (P74-s2work)\color{black}. Participants also hope for automation within their specialized work domains, like \emph{```status updates' from project management activities''} (P332-s1personal\color{black}) and \emph{``hardware design''} (P83-s1nouse\color{black}). 

In Survey2, participants had further ideas for automation, like \emph{``tracking purchase order data more efficiently''} (P89-s2work), \emph{``delivering reports via email or upload files to Box folder regularly on a schedule''} (P101-s2work), and \emph{``expense reports, time reports, booking leave, client contacts''} (P107-s2work). 
\color{black}
We can imagine ways that LLMs can enable these kinds of scenarios, like a user inputting a description in natural language that a system then translates to a hardware design. Yet, the systems would likely be more complex than one LLM taking an input and providing an output.
Further, these automation tasks are not solitary disconnected tasks, but would fit into larger workflows.
LLM-based systems are moving towards supporting automation, especially with tool calling capabilities and LLM reasoning for complex tasks~\cite{yao2023reactsynergizingreasoningacting}, though these capabilities are still being researched and few are available to knowledge workers at this point. \color{black}

\subsection{RQ3: How does LLM usage fit into larger and collaborative workflows?}

Because work often involves many inter-connected, collaborative processes, we wanted to understand how current usage of LLMs was part of participants' workflows. We asked work-use  (53) \color{black} and personal-use  (60) participants in Survey1 and work-use participants (37) in Survey2 \color{black} about whether their use of LLMs was part of larger workflows and whether those workflows were collaborative. While we did not specifically ask about how future LLM tasks fit into workflows to keep the survey a reasonable length, we found that participants described multiple related assistance types or tasks that would involve combinations of LLM assistance. 

\subsubsection{Current Workflows}
We asked participants if their LLM tasks were part of larger workflows (yes/no), what the workflows were, if those larger workflows were collaborative (yes/no), who else was involved in the workflows (Survey1 only), challenges in collaborative or complex workflows (Survey2 only), \color{black} and how they did the tasks in the past. 
More than half of work-use LLM tasks (60\%) were part of larger tasks, compared to only 30\% of personal-use LLM tasks  in Survey1. In Survey2, only 42\% of work-use tasks were part of larger workflows. Our relatively smaller sample size for this question in Survey2 may account for the difference. Participants may have also become more independent in their use of LLMs, feeling more confident to use LLMs in more solitary tasks, rather than using them in more collaborative spaces. \color{black}
Further, work-use workflows were often collaborative (69\%) compared to personal-use LLM workflows (52\%)  in Survey1. In Survey2, 70\% of larger workflows were collaborative for work tasks. \color{black}
However, few participants had used LLMs for multiple parts of a workflow so far (11\% work-use, 5\% personal-use  in Survey1\color{black}). 
Participants described workflows from a high level like \emph{``Software development: expanding on designs, code generation and test case generation''} (P345-s1work\color{black}) or \emph{``the analysis process of user research''} (P248-s1work\color{black}), to more specific like \emph{``debugging a commit''} (P360-s1work\color{black}). Participants also acknowledged that the text they were asking for help with was part of larger works, like a presentation (P92-s1work\color{black}) or creating a report (P327-s1work\color{black}). 

In terms of collaborative efforts, P282-s1work \color{black} sums up the nature of other people involved in these processes: \emph{``There are many.''} Participants described multiple people involved in the workflows, naming their teams as well as multiple other teams that they work with. For example, P198-s1work \color{black} wrote: \emph{``Designers, PMs, devs''}.
How participants completed their tasks pre-LLM also provides insight into their workflows. Prior to using an LLM, Survey1 \color{black}participants primarily completed their tasks themselves manually (s1work-use\color{black}: 68\%, s1personal-use\color{black}: 60\%). Participants had also used other tools (s1work-use\color{black}: 17\%, s1personal-use\color{black}: 18\%), had help from others (s1work-use\color{black}: 8\%, s1personal-use\color{black}: 3\%), or had not done the task prior to using an LLM for it (s1work-use\color{black}: 8\%, s1personal-use\color{black}: 15\%). Primarily, workers have used LLMs as support for tasks they normally do themselves. 

 In Survey2, we asked participants about challenges they encountered in using LLMs collaboratively or in larger workflows. While we didn't have enough data to perform a full analysis, we report some anecdotes as starting points for further study. Two participants talked about lack of repeatability. P85-s2work wrote: \emph{``the same prompt used by different people can generate differing responses''} and P95-s2work wrote \emph{``Repeatability is non-existent. Artifacts that may pop up once, just go away if someone else is using the same inputs.''}
P88-s2work wrote about not trusting co-workers who use LLMs, as their work may not be as high quality: \emph{``I fundamentally don't trust the work of people who I know use LLMs for their work, because I know how likely it is the work is poor quality.''}
Finally, in terms of larger workflows, P44-s2work wrote: \emph{``Integration is challenging when working with LLM''}. 
Yet, nine participants wrote that they didn't have any challenges in collaborative or larger workflows including LLMs.
\color{black}

\subsubsection{Future Workflows}
While nearly all current LLM usages were discrete tasks, some participants described future visions of LLM usage that span across workflows. We selected three examples of workflows from the set of future tasks that illustrate how multiple types of LLM support could be integrated together: 1) Communications, 2) Information Management, and 3) Role-based.

In the communication domain, there are a variety of tasks that a human assistant might help a worker with, such as drafting emails (creation-artifact), improving emails and messages (advice-improve), summarizing meeting notes (information-summarize), and scheduling meetings (automation). 
Participants envision a future where an LLM-based system supports them throughout many of these tasks. For example, P186-s1work \color{black} said they would like an LLM to: \emph{``handle my calendar, to take good notes and summarize from meeting instead of me having to do... to summarize my emails and important parts of it so I don't get drowned in emails.''} 
Some participants also hoped for a system with more specialization, such as one that could make recommendations based on conversations (information-analyze) and suggest target due dates (advice-guidance) (P42-s1work\color{black}).
However, workers also have concerns about turning over so much of a workflow to a system. For example, P95-s1personal \color{black} said:
\emph{``While in theory it sounds great to use LLMs to put together emails, etc., in reality, the amount of work that would be required to edit and tailor those emails means time savings in my position would likely be limited.''} 
Not only does the content need to be correct, but the style of these types of communications are often critical in a business context.

Participants had several ideas for ways LLMs could help them manage their information, such as \emph{``Improve ability to quickly find information across platforms like [anonymized application names] and local files. This is currently a significant source of inefficiency.''} (P331-s1tryit\color{black}), or dashboards that \emph{``centralize, prioritizes information from multiple sources so I can execute more efficiently and effectively''} (P47-s1personal\color{black}) and provide \emph{``status updates, milestone dates, etc''} (P98-s1personal\color{black}). These tasks include searching through information, summarizing and analyzing important parts of information, as well as automation to keep track of data across multiple sources. One participant hoped that an LLM could also give them access to their colleague's information (information-search), so that they could \emph{``ask it questions when other developers are gone for the day or have left the company.''} (P39-s1work\color{black}). An LLM-based information dashboard would likely involve many of the information tasks, like search, learn, summarize and analyze, and could also include guidance, validation, as well as automation.

Finally, participants also envision tools that can provide support with many of their specialized role-based tasks, both technical and non-technical. 
One participant listed a variety of tasks they would like for an LLM to help with: \emph{``Refactor entire codebases, for example, to switch the underlying framework or change a dependency to a different one all over the code base. Analyze GitHub issues, bug reports and create first draft pull requests that address the issue. Do code reviews that analyze the pull request description to check if it's properly worded, and to see if the description matches the actual changes in the files.''} (P135-s1work\color{black}). Within this set of tasks, a tool would need to be able to do complex code generation (creation-artifact), generate pull requests based on issues (creation-artifact, information-analyze), and check whether pull requests are valid (advice-validation). 
Participants also described wanting help with a variety of non-technical role-based tasks, such as UX Research tasks, like analysis (information-analysis), synthesis (information-analysis, information-summarization), and presentation of results (creation-artifact), though \emph{``always in a passenger seat''} (P110-s1personal\color{black}). In these descriptions, participants would like an LLM to have similar expertise as they need to have in their own role. However, because these tasks and workflows are critical outcomes of workers' jobs, they would expect to oversee the results. Participants also envision the LLMs integrated throughout their tooling and workflows to be able to support their tasks.

\section{Discussion}
We discuss: 1) our set of LLM usages in the context of prior work  and the feasibility of desired LLM uses \color{black}, 2) implications for future LLM support for knowledge workers, and 3) limitations. 

\subsection{Current and Future LLM Use Cases}
One primary result of this work is an analysis of the ways knowledge workers currently use LLMs and want to use LLMs in the future. Our findings both support and expand upon previous findings. Researchers have described partial sets of our LLM uses in current general use of LLMs~\cite{skjuve2023user, taecharungroj2023can, skjuve4376834people}, as well as particular contexts like writing~\cite{gmeiner2023dimensions}, programming~\cite{ross2023programmer}, healthcare~\cite{zheng2023innovating}, and education~\cite{yan2023practical}. Our usage themes and sub-themes capture LLM usages across multiple prior studies in various domains (see Table~\ref{tab:discussion}), providing a detailed and cohesive view of LLM use for knowledge work. 
The closest work to ours, which covered interviews of knowledge workers outlined a set of knowledge work tasks that LLMs can support that align with 6 of the 10 types of LLM usage that we describe~\cite{ritala2023transforming}. Our results indicate that workers of a variety of roles, including non-technical ones, already use LLMs for generation of work documents, ideas, learning and finding information, and improving their writing. In the future, workers hope to be able to use LLMs to capture insights about their own data as well and automate their tasks. Both current and future tasks require that the LLM has certain abilities in order to generate valuable content or provide correct information. Further, we found that use of LLMs at work is often within larger collaborative workflows. Finally, both current and future use of LLMs at work requires oversight. Our study provides a foundation for future work designing tools to support workers using LLMs. 
\begin{table}[b]
\caption{Support and expansion of prior work with our findings }~\label{tab:discussion}
\small
\begin{tabular}{lll}
\toprule
Theme & Sub-Theme & Prior work that includes this type of LLM task \\
\hline
\multirow{2}{*}{Creation} & Artifact & work~\cite{ritala2023transforming}, writing~\cite{gmeiner2023dimensions}, education~\cite{yan2023practical},  general~\cite{skjuve2023user, taecharungroj2023can, skjuve4376834people}\\
& Idea &  work~\cite{ritala2023transforming}, programming~\cite{ross2023programmer}, general~\cite{skjuve2023user,taecharungroj2023can,skjuve4376834people}\\ \hline
\multirow{2}{*}{Information} & Search & work~\cite{ritala2023transforming}, programming~\cite{ross2023programmer}, general~\cite{skjuve2023user,taecharungroj2023can,skjuve4376834people}\\
& Learn & work~\cite{ritala2023transforming}, programming~\cite{ross2023programmer}, general~\cite{skjuve2023user,taecharungroj2023can,skjuve4376834people}\\
& Summarize &   writing~\cite{gmeiner2023dimensions} \\
& Analyze & education~\cite{yan2023practical}, healthcare~\cite{zheng2023innovating}\\ \hline
\multirow{3}{*}{Advice} & Improve & work~\cite{ritala2023transforming}, programming~\cite{ross2023programmer}, writing~\cite{gmeiner2023dimensions}\\
& Guidance &  education~\cite{yan2023practical}, healthcare~\cite{zheng2023innovating}\\
& Validation & education~\cite{yan2023practical} \\ \hline

Automation& Automation & work~\cite{ritala2023transforming} \\

\bottomrule
\end{tabular}
\end{table}

One question that arises is whether participants described future uses of LLMs that are actually feasible. While some of the uses cases described have not yet been created, one of the main potential questions may be about the integration of LLMs into larger workflows, as this presents many challenges. Yet, as one example, Microsoft's Copilot for 365 was announced in September 2023~\footnote{https://www.microsoft.com/en-us/microsoft-365/blog/2023/09/21/announcing-microsoft-365-copilot-general-availability-and-microsoft-365-chat/} (after Survey1 had concluded) and includes significant user data and application integration. This tooling also includes support in software like Excel for tasks like analyzing information. Tools like IBM's WatsonX Orchestrate~\footnote{https://www.ibm.com/products/watsonx-orchestrate/skills} provide ways to create automation pipelines using LLMs. LLMs are also being integrated into tools used in knowledge work, like Airtable~\footnote{https://www.airtable.com/platform/ai}, Mural~\footnote{https://www.mural.co/blog/announcing-mural-ai}, and Figma~\footnote{https://www.figma.com/ai/}. These tools may not yet fully satisfy the needs of users, but indicate feasibility of participants' desired tasks and that systems are moving towards LLM capabilities for these tasks. Further, LLM-based systems are moving toward more complex functionalities like tool-calling to integrate external information and capabilities, as well as LLM-based reasoning~\cite{yao2023reactsynergizingreasoningacting}, which can support more complex tasks. 
\color{black}

\subsection{Design Implications and Challenges for LLM-based Tools for Knowledge Workers}
We present three key outcomes of this research: workers need LLM-based tools that support their specialized skills, support integrated and data-heavy workflows, and support worker-LLM collaboration. We describe challenges and opportunities for future work in these areas. 

\subsubsection{Support specialized skills}
Across the tasks users described both for current and future LLM usages, specialized skills are critical. 
For example, generation of code or text for a business context require certain levels of skill and professionalism. Users seeking information or to learn about a topic need to be sure that the model has been trained on the necessary information. There are two implications of this: 1) workers need to be able to determine if an LLM or LLM-based systems has been trained on the necessary information to be helpful for their task, and 2) workers may benefit from being able to customize their LLMs or LLM-based tools with role or company-specific information. 

Transparency may help workers determine whether an LLM-based tool has critical knowledge. Documentation about LLMs sometimes provides transparency about the training data. However, this transparency is often missing in proprietary models or this may not be detailed enough for a user to know whether the LLM has the expertise to be able to complete specialized work tasks~\cite{liao2023ai}. Reporting frameworks have been developed for AI systems, like FactSheets~\cite{arnold2019factsheets} or model cards~\cite{mitchell2019model}, but users likely need more than static documentation~\cite{liao2023designerly,crisan2022interactive}. If the system is performing a task that the user is an expert in, they may be able to perform an evaluation to determine if the outputs are sufficient. However, if they are using an LLM to learn or find information, they may not be able to evaluate that system. Future work could explore how workers, especially end-users, are currently evaluating LLM capabilities, as our work suggests that they are doing some light-weight evaluation. Additionally, one potential direction in this space is to leverage communities of workers of varying skill levels who can provide social transparency~\cite{ehsan2021expanding} for LLM systems in work contexts. 

Even if an LLM-based system has many or most of the capabilities necessary to support a particular job role, there may be specific details or knowledge for a worker or company. For example, we had one worker who found that ChatGPT provided information about Kubernetes at the correct level and this was important, but that same information would not necessarily be at the appropriate level for another worker. Another worker needed to edit draft responses in `brand voice' for the particular company. While a user could provide some of this context in a conversational agent, this might not work if the customized content is more extensive. Further, many LLM-based systems are not conversational. Early research has begun to investigate personalized text generation especially on the algorithmic side~\cite{li2023teach, eapenpersonalization}, but future work could explore how to enable workers to customize LLM-based tools for their specialized contexts. 

\subsubsection{Support Integrated Workflows and Data}
Workers described desired future LLM usage in which LLMs are integrated into their workflows and leverage their own data. During the time period we collected responses, participants in our population did not have access to integrated LLM tools that have begun to be released, such as Microsoft's CoPilot\footnote{https://blogs.microsoft.com/blog/2023/03/16/introducing-microsoft-365-copilot-your-copilot-for-work/} for tools like Word, PowerPoint, Excel, and email or the AI companion within Zoom\footnote{https://www.zoom.com/en/blog/zoom-ai-companion/}. These tools may begin to address some of the needs and desires of workers, such as presentation and spreadsheet building, as these AI tools will have access to the user's data. 
Other scenarios, like tools for supporting sharing of complex information sources or integrating data from multiple tools or sources, may not fit as cleanly into AI tools within existing work tools. Instead, users are looking to be able to integrate multiple sources of information depending on their context. One challenge in this space is that workers have individual and contextual workflows~\cite{kogan2006ethnographic}. End-user programming may be one way to support these needs, such as `if this then that'~\footnote{https://ifttt.com/}, which could allow users to import the data they need and set up their own flows with LLM components. Some research has begun to explore visual programming environments for LLMs components like PromptChainer~\cite{wu2022promptchainer} and ChainForge~\cite{arawjo2023chainforge}, but workflow and data automation in the workplace may also need non-LLM components and functionalities, adding complexity.

\subsubsection{Support Worker-LLM collaboration}
Across both current and future descriptions of LLM usage for work, participants described using LLMs for ideas, drafting, and as an assistant. However, due to the importance of work outputs and the unpredictable nature of generative AI, workers need to oversee LLM outputs. Overseeing the outputs could range from a quick check to extensive editing, leading to extra work regardless of whether it is tedious or a significant effort. In some ways, the task of overseeing an LLM output has overlap with the task of evaluating LLM output, such as during prompt engineering and system development. However, an end-user may be more willing and able to modify the LLM output than modify their input to the LLM to try to improve the output. We suggest that future systems for knowledge workers provide support for workers to oversee and modify LLM outputs in ways that minimize effort while ensuring the necessary quality of work. \color{black} While our data indicate that most current uses of LLMs are individual work, the output of worker-LLM collaboration indirectly impacts team collaboration. We found that using LLMs in larger workflows may raise issues with co-workers trusting less in the quality of the work of other team members if they are using AI. Future systems supporting knowledge workers may want to address these issues, such as by incorporating methods for AI attribution \cite{he2025contributionsdeservecreditperceptions}.
\color{black}

\subsubsection{Support Collaborative Knowledge Work}
\color{black}
Our results align with prior work, indicating that knowledge work is often complex and collaborative~\cite{iivari1999knowledge}. However, many existing generative AI tools for knowledge work are not inherently collaborative, limiting our ability to deeply probe into collaborative work with the use of LLM tools at this time. As LLMs and the technologies around them mature, we expect their roles to evolve from assistants to autonomous AI agents that can collaborate more directly on a team. 

Participants described several possible concerns for future LLM-based tools for collaborative work, such as repeatability, trust, and integration. Participants also described a diverse set of future tasks they would use LLMs for, with a range of collaboration types. The type of collaboration involved is likely to impact users' needs and challenges with regard to LLM-based support. 
Some types of tasks, such as creating an idea or brainstorming, lend themselves to LLM integration directly into the collaborative context. Recent research has shown the potential of LLM integration to benefit group brainstorming~\cite{muller2024group, shaer2024toward}. In this kind of context, concerns like repeatability and trust may be less salient than the group's ability to effectively interact with, control, and integrate the LLM tool within their workflow. 
Participants also talked about wanting LLMs to support their communications workflows, such as writing emails and scheduling meetings. These workflows involve multiple people, but the collaboration is often more asynchronous in nature than a group brainstorming session. Effective integration with the existing tools and processes would likely be critical in this context, as well as consistency and reliability of a tool's actions. 
Our findings suggest that future work should investigate the needs of knowledge workers within the context of the type of  collaboration in order to best support their use of LLMs.

\color{black}

\subsection{Limitations}
We have several limitations in our methods: 1) our population, 2) not specifying or capturing the LLMs used, and 3) our survey formats. \color{black}
We completed this survey internally at one large international technology company. While we recruited broadly across the company, including non-technical roles, many of our participants (S1: \color{black}45\%, S2: 48\%\color{black}) did have a technical role. Our population is also heavily North American, Western European, and Asian. 
Further, the particular company we recruited from had policies in place limiting use of some LLMs for work purposes due to privacy concerns. 
 A few participants (17/216 in Survey1 and 9/107 in Survey2) said they don't know much about AI, which may have limited their ability to speculate about how they would use LLMs in the future at work. Yet, 151/216 of Survey1 participants and 71/107 Survey2 participants had at least tried consumer tools. Further, all participants in both surveys reported that their work is at least somewhat related to AI, which makes sense as a large focus of the company is on AI. The company also conducted a large scale education campaign on generative AI between Survey1 and Survey2. \color{black}
We do not capture the particular LLMs used, in order to provide anonymity to our participants and the systems used.
In order to capture broad usage of LLMs, we did not specify usage of one particular LLM. Due to these choices, we are unable to link uses and challenges to particular LLMs or systems. However, we believe this lends generalizability to our results, as we do not narrow to one LLM or system. 

Finally, our survey results limit the depth of detail we obtained about the use cases. However, we believe a survey is an important step in understanding the breadth of current and desired uses for LLMs for knowledge workers in a technology company. Our two surveys also had different formats, as the purpose of the first survey was to understand the types of tasks and the second survey had participants select from those types in order to better understand frequencies of tasks. Future work could continue to replicate our survey to capture longer-term trends, further investigate high-impact use cases, and dig deeper into the needs for LLM tools in collaborative contexts. \color{black} Due to the timing of these surveys, the tools available to knowledge workers were primarily non-collaborative, limiting the ability of participants to comment on their use of and needs for collaborative tools. As tools become more available for larger and more collaborative workflows, future work will be better able to study and address user needs in these contexts.
\color{black}

\section{Conclusion}
In this paper, we present two surveys from 2023 and 2024 \color{black} of knowledge workers' current uses and future desired uses of LLMs for four categories of LLM adoption.  In the first survey, we found that \color{black} about a quarter of participants used LLMs at work (work-use), primarily requesting \textit{creation} of templates or starting points of code or text that they then planned to modify. Outside of work (personal-use), participants often asked LLMs for \textit{ideas} on a variety of specialized domains, while those who had only experimented with LLMs (tried-it) often wanted to see how an LLM would answer questions. Participants' future visions of using LLMs involved \textit{automation} and having the LLMs integrated into their data and workflows. 

In the second survey, we found an increase in use of LLMs at work \color{black}across use types, indicating broad adoption of LLMs into knowledge work, \color{black} with 34.6\% of participants using LLMs for their job roles (compared to 24.5\% in the first survey), and only 8.4\% not using them at all (compared to 20.8\% not having used them in the first survey). Our second survey further validated the categories of tasks we found and provides an updated view into the types of tasks knowledge workers are currently using LLMs for (creation, searching, learning, and summarizing) and which tasks they would like to use LLMs for in the future (analyzing information, advice, and automation). \color{black} The current use and desired future use of LLMs in automation may be indicative of a shift from individual tool use toward LLMs being more integrated as autonomous agents into collaborative teams.
\color{black}
We contribute to discussion of adoption of LLMs and designing LLM-based tools. This is a pivotal moment in the evolution of LLMs, with rapid technological advancement, where nuanced analysis into adoption by users is critical to the designers and builders of these tools. Our survey provides insights into how to capture adoption of LLMs by knowledge workers, a process which has only just begun.


\bibliographystyle{ACM-Reference-Format}
\bibliography{LLMUseCaseSurvey-arxiv/0_main}


\begin{thebibliography}{88}


\ifx \showCODEN    \undefined \def \showCODEN     #1{\unskip}     \fi
\ifx \showDOI      \undefined \def \showDOI       #1{#1}\fi
\ifx \showISBNx    \undefined \def \showISBNx     #1{\unskip}     \fi
\ifx \showISBNxiii \undefined \def \showISBNxiii  #1{\unskip}     \fi
\ifx \showISSN     \undefined \def \showISSN      #1{\unskip}     \fi
\ifx \showLCCN     \undefined \def \showLCCN      #1{\unskip}     \fi
\ifx \shownote     \undefined \def \shownote      #1{#1}          \fi
\ifx \showarticletitle \undefined \def \showarticletitle #1{#1}   \fi
\ifx \showURL      \undefined \def \showURL       {\relax}        \fi
\providecommand\bibfield[2]{#2}
\providecommand\bibinfo[2]{#2}
\providecommand\natexlab[1]{#1}
\providecommand\showeprint[2][]{arXiv:#2}

\bibitem[Bar({[n.\,d.]})]%
        {Bard}
 \bibinfo{year}{[n.\,d.]}\natexlab{}.
\newblock \bibinfo{title}{Bard- Chat Based AI Tool from Google, Powered by PaLM
  2}.
\newblock \bibinfo{howpublished}{\url{https://bard.google.com/}}.
\newblock
\newblock
\shownote{Accessed: 2023-09-02}.


\bibitem[Cha({[n.\,d.]})]%
        {ChatGPT}
 \bibinfo{year}{[n.\,d.]}\natexlab{}.
\newblock \bibinfo{title}{ChatGPT}.
\newblock \bibinfo{howpublished}{\url{https://chat.openai.com/}}.
\newblock
\newblock
\shownote{Accessed: 2023-07-28}.


\bibitem[CoP({[n.\,d.]})]%
        {CoPilot}
 \bibinfo{year}{[n.\,d.]}\natexlab{}.
\newblock \bibinfo{title}{GitHub Copilot - Your AI pair programmer}.
\newblock \bibinfo{howpublished}{\url{https://github.com/features/copilot}}.
\newblock
\newblock
\shownote{Accessed: 2023-07-28}.


\bibitem[Cha(2023)]%
        {ChatGPTCatchingOn}
 \bibinfo{year}{2023}\natexlab{}.
\newblock \bibinfo{title}{How ChatGPT Is Catching On in America}.
\newblock
  \bibinfo{howpublished}{\url{https://wordfinder.yourdictionary.com/blog/how-chatgpt-is-catching-on-in-america/}}.
\newblock
\newblock
\shownote{Accessed: 2023-08-21}.


\bibitem[McK(2023)]%
        {McKinsey}
 \bibinfo{year}{2023}\natexlab{}.
\newblock \bibinfo{title}{Jobs lost, jobs gained: What the future of work will
  mean for jobs, skills, and wages}.
\newblock
  \bibinfo{howpublished}{\url{https://www.mckinsey.com/featured-insights/future-of-work/jobs-lost-jobs-gained-what-the-future-of-work-will-mean-for-jobs-skills-and-wages}}.
\newblock
\newblock
\shownote{Accessed: 2023-07-28}.


\bibitem[Agossah et~al\mbox{.}(2023)]%
        {agossah2023llmbased}
\bibfield{author}{\bibinfo{person}{Alexandre Agossah},
  \bibinfo{person}{Frédérique Krupa}, \bibinfo{person}{Matthieu Perreira~Da
  Silva}, {and} \bibinfo{person}{Patrick~Le Callet}.}
  \bibinfo{year}{2023}\natexlab{}.
\newblock \bibinfo{title}{LLM-based Interaction for Content Generation: A Case
  Study on the Perception of Employees in an IT department}.
\newblock
\newblock


\bibitem[Arawjo et~al\mbox{.}(2023)]%
        {arawjo2023chainforge}
\bibfield{author}{\bibinfo{person}{Ian Arawjo}, \bibinfo{person}{Priyan
  Vaithilingam}, \bibinfo{person}{Martin Wattenberg}, {and}
  \bibinfo{person}{Elena Glassman}.} \bibinfo{year}{2023}\natexlab{}.
\newblock \showarticletitle{ChainForge: An open-source visual programming
  environment for prompt engineering}. In \bibinfo{booktitle}{\emph{Adjunct
  Proceedings of the 36th Annual ACM Symposium on User Interface Software and
  Technology}}. \bibinfo{pages}{1--3}.
\newblock


\bibitem[Arnold et~al\mbox{.}(2019)]%
        {arnold2019factsheets}
\bibfield{author}{\bibinfo{person}{Matthew Arnold}, \bibinfo{person}{Rachel~KE
  Bellamy}, \bibinfo{person}{Michael Hind}, \bibinfo{person}{Stephanie Houde},
  \bibinfo{person}{Sameep Mehta}, \bibinfo{person}{Aleksandra Mojsilovi{\'c}},
  \bibinfo{person}{Ravi Nair}, \bibinfo{person}{K~Natesan Ramamurthy},
  \bibinfo{person}{Alexandra Olteanu}, \bibinfo{person}{David Piorkowski},
  {et~al\mbox{.}}} \bibinfo{year}{2019}\natexlab{}.
\newblock \showarticletitle{FactSheets: Increasing trust in AI services through
  supplier's declarations of conformity}.
\newblock \bibinfo{journal}{\emph{IBM Journal of Research and Development}}
  \bibinfo{volume}{63}, \bibinfo{number}{4/5} (\bibinfo{year}{2019}),
  \bibinfo{pages}{6--1}.
\newblock


\bibitem[Bannon et~al\mbox{.}(1983)]%
        {bannon1983evaluation}
\bibfield{author}{\bibinfo{person}{Liam Bannon}, \bibinfo{person}{Allen
  Cypher}, \bibinfo{person}{Steven Greenspan}, {and} \bibinfo{person}{Melissa~L
  Monty}.} \bibinfo{year}{1983}\natexlab{}.
\newblock \showarticletitle{Evaluation and analysis of users' activity
  organization}. In \bibinfo{booktitle}{\emph{Proceedings of the SIGCHI
  conference on Human Factors in Computing Systems}}. \bibinfo{pages}{54--57}.
\newblock


\bibitem[Berg and Gmyrek(2023)]%
        {berg2023automation}
\bibfield{author}{\bibinfo{person}{Janine Berg} {and} \bibinfo{person}{Pawel
  Gmyrek}.} \bibinfo{year}{2023}\natexlab{}.
\newblock \showarticletitle{Automation Hits the Knowledge Worker: ChatGPT and
  the Future of Work}. In \bibinfo{booktitle}{\emph{UN Multi-Stakeholder Forum
  on Science, Technology and Innovation for the SDGs (STI Forum)}}.
\newblock


\bibitem[Bloomberg(2023)]%
        {IBM-Bloomberg}
Bloomberg \bibinfo{year}{2023}\natexlab{}.
\newblock \bibinfo{title}{IBM to Pause Hiring for Jobs that AI Could Do}.
\newblock \bibinfo{howpublished}{Bloomberg}.
\newblock
\urldef\tempurl%
\url{https://www.bloomberg.com/news/articles/2023-05-01/ibm-to-pause-hiring-for-back-office-jobs-that-ai-could-kill#xj4y7vzkg}
\showURL{%
\tempurl}
\newblock
\shownote{Accessed: 2023-07-28}.


\bibitem[Bonsu and Baffour-Koduah(2023)]%
        {bonsu2023consumers}
\bibfield{author}{\bibinfo{person}{Emmanuel Bonsu} {and}
  \bibinfo{person}{Daniel Baffour-Koduah}.} \bibinfo{year}{2023}\natexlab{}.
\newblock \showarticletitle{From the consumers’ side: Determining students’
  perception and intention to use ChatGPTin ghanaian higher education}.
\newblock \bibinfo{journal}{\emph{Available at SSRN 4387107}}
  (\bibinfo{year}{2023}).
\newblock


\bibitem[Braun and Clarke(2006)]%
        {braun2006using}
\bibfield{author}{\bibinfo{person}{Virginia Braun} {and}
  \bibinfo{person}{Victoria Clarke}.} \bibinfo{year}{2006}\natexlab{}.
\newblock \showarticletitle{Using thematic analysis in psychology}.
\newblock \bibinfo{journal}{\emph{Qualitative research in psychology}}
  \bibinfo{volume}{3}, \bibinfo{number}{2} (\bibinfo{year}{2006}),
  \bibinfo{pages}{77--101}.
\newblock


\bibitem[Braun and Clarke(2021)]%
        {braun2021one}
\bibfield{author}{\bibinfo{person}{Virginia Braun} {and}
  \bibinfo{person}{Victoria Clarke}.} \bibinfo{year}{2021}\natexlab{}.
\newblock \showarticletitle{One size fits all? What counts as quality practice
  in (reflexive) thematic analysis?}
\newblock \bibinfo{journal}{\emph{Qualitative research in psychology}}
  \bibinfo{volume}{18}, \bibinfo{number}{3} (\bibinfo{year}{2021}),
  \bibinfo{pages}{328--352}.
\newblock


\bibitem[Burgess et~al\mbox{.}(2023)]%
        {burgess2023healthcare}
\bibfield{author}{\bibinfo{person}{Eleanor~R Burgess}, \bibinfo{person}{Ivana
  Jankovic}, \bibinfo{person}{Melissa Austin}, \bibinfo{person}{Nancy Cai},
  \bibinfo{person}{Adela Kapu{\'s}ci{\'n}ska}, \bibinfo{person}{Suzanne
  Currie}, \bibinfo{person}{J~Marc Overhage}, \bibinfo{person}{Erika~S Poole},
  {and} \bibinfo{person}{Jofish Kaye}.} \bibinfo{year}{2023}\natexlab{}.
\newblock \showarticletitle{Healthcare AI Treatment Decision Support: Design
  Principles to Enhance Clinician Adoption and Trust}. In
  \bibinfo{booktitle}{\emph{Proceedings of the 2023 CHI Conference on Human
  Factors in Computing Systems}}. \bibinfo{pages}{1--19}.
\newblock


\bibitem[Carvalho and Ivanov(2023)]%
        {carvalho2023chatGPT}
\bibfield{author}{\bibinfo{person}{Inês Carvalho} {and}
  \bibinfo{person}{Stanislav Ivanov}.} \bibinfo{year}{2023}\natexlab{}.
\newblock \bibinfo{title}{ChatGPT for tourism: applications, benefits and
  risks}.
\newblock
\newblock


\bibitem[Chan and Hu(2023)]%
        {chan2023students}
\bibfield{author}{\bibinfo{person}{Cecilia Ka~Yuk Chan} {and}
  \bibinfo{person}{Wenjie Hu}.} \bibinfo{year}{2023}\natexlab{}.
\newblock \bibinfo{title}{Students' Voices on Generative AI: Perceptions,
  Benefits, and Challenges in Higher Education}.
\newblock
\newblock


\bibitem[Chilana et~al\mbox{.}(2015)]%
        {chilana2015user}
\bibfield{author}{\bibinfo{person}{Parmit~K Chilana}, \bibinfo{person}{Amy~J
  Ko}, {and} \bibinfo{person}{Jacob Wobbrock}.}
  \bibinfo{year}{2015}\natexlab{}.
\newblock \showarticletitle{From user-centered to adoption-centered design: a
  case study of an HCI research innovation becoming a product}. In
  \bibinfo{booktitle}{\emph{Proceedings of the 33rd Annual ACM Conference on
  Human Factors in Computing Systems}}. \bibinfo{pages}{1749--1758}.
\newblock


\bibitem[Crisan et~al\mbox{.}(2022)]%
        {crisan2022interactive}
\bibfield{author}{\bibinfo{person}{Anamaria Crisan}, \bibinfo{person}{Margaret
  Drouhard}, \bibinfo{person}{Jesse Vig}, {and} \bibinfo{person}{Nazneen
  Rajani}.} \bibinfo{year}{2022}\natexlab{}.
\newblock \showarticletitle{Interactive model cards: A human-centered approach
  to model documentation}. In \bibinfo{booktitle}{\emph{Proceedings of the 2022
  ACM Conference on Fairness, Accountability, and Transparency}}.
  \bibinfo{pages}{427--439}.
\newblock


\bibitem[Dabbish and Kraut(2006)]%
        {dabbish2006email}
\bibfield{author}{\bibinfo{person}{Laura~A Dabbish} {and}
  \bibinfo{person}{Robert~E Kraut}.} \bibinfo{year}{2006}\natexlab{}.
\newblock \showarticletitle{Email overload at work: An analysis of factors
  associated with email strain}. In \bibinfo{booktitle}{\emph{Proceedings of
  the 2006 20th anniversary conference on Computer supported cooperative
  work}}. \bibinfo{pages}{431--440}.
\newblock


\bibitem[Dahlkemper et~al\mbox{.}(2023)]%
        {dahlkemper2023physics}
\bibfield{author}{\bibinfo{person}{Merten~Nikolay Dahlkemper},
  \bibinfo{person}{Simon~Zacharias Lahme}, {and} \bibinfo{person}{Pascal
  Klein}.} \bibinfo{year}{2023}\natexlab{}.
\newblock \bibinfo{title}{How do physics students evaluate artificial
  intelligence responses on comprehension questions? A study on the perceived
  scientific accuracy and linguistic quality}.
\newblock
\newblock


\bibitem[Das~Swain et~al\mbox{.}(2023)]%
        {10.1145/3544548.3581326}
\bibfield{author}{\bibinfo{person}{Vedant Das~Swain}, \bibinfo{person}{Javier
  Hernandez}, \bibinfo{person}{Brian Houck}, \bibinfo{person}{Koustuv Saha},
  \bibinfo{person}{Jina Suh}, \bibinfo{person}{Ahad Chaudhry},
  \bibinfo{person}{Tenny Cho}, \bibinfo{person}{Wendy Guo},
  \bibinfo{person}{Shamsi Iqbal}, {and} \bibinfo{person}{Mary~P Czerwinski}.}
  \bibinfo{year}{2023}\natexlab{}.
\newblock \showarticletitle{Focused Time Saves Nine: Evaluating
  Computer–Assisted Protected Time for Hybrid Information Work}. In
  \bibinfo{booktitle}{\emph{Proceedings of the 2023 CHI Conference on Human
  Factors in Computing Systems}} (Hamburg, Germany) \emph{(\bibinfo{series}{CHI
  '23})}. \bibinfo{publisher}{Association for Computing Machinery},
  \bibinfo{address}{New York, NY, USA}, Article \bibinfo{articleno}{857},
  \bibinfo{numpages}{18}~pages.
\newblock
\showISBNx{9781450394215}
\urldef\tempurl%
\url{https://doi.org/10.1145/3544548.3581326}
\showDOI{\tempurl}


\bibitem[Davenport(2005)]%
        {davenport2005thinking}
\bibfield{author}{\bibinfo{person}{Thomas~H Davenport}.}
  \bibinfo{year}{2005}\natexlab{}.
\newblock \bibinfo{booktitle}{\emph{Thinking for a living: how to get better
  performances and results from knowledge workers}}.
\newblock \bibinfo{publisher}{Harvard Business Press}.
\newblock


\bibitem[DEM{\.I}R and DEM{\.I}R(2023)]%
        {demir2023professionals}
\bibfield{author}{\bibinfo{person}{{\c{S}}irvan~{\c{S}}en DEM{\.I}R} {and}
  \bibinfo{person}{Mahmut DEM{\.I}R}.} \bibinfo{year}{2023}\natexlab{}.
\newblock \showarticletitle{Professionals' perspectives on ChatGPT in the
  tourism industry: Does it inspire awe or concern?}
\newblock \bibinfo{journal}{\emph{Journal of Tourism Theory and Research}}
  \bibinfo{volume}{9}, \bibinfo{number}{2} (\bibinfo{year}{2023}),
  \bibinfo{pages}{61--76}.
\newblock


\bibitem[Di~Ciccio and Mecella(2013)]%
        {di2013mining}
\bibfield{author}{\bibinfo{person}{Claudio Di~Ciccio} {and}
  \bibinfo{person}{Massimo Mecella}.} \bibinfo{year}{2013}\natexlab{}.
\newblock \showarticletitle{Mining artful processes from knowledge workers'
  emails}.
\newblock \bibinfo{journal}{\emph{IEEE Internet Computing}}
  \bibinfo{volume}{17}, \bibinfo{number}{5} (\bibinfo{year}{2013}),
  \bibinfo{pages}{10--20}.
\newblock


\bibitem[Drebert et~al\mbox{.}(2023)]%
        {drebert2023influence}
\bibfield{author}{\bibinfo{person}{Judith Drebert}, \bibinfo{person}{Sid
  Suntrayuth}, {and} \bibinfo{person}{Stephan Böhm}.}
  \bibinfo{year}{2023}\natexlab{}.
\newblock \bibinfo{title}{The influence of perceived transparency on the
  acceptance of work task automation: an example of recruiting chatbots in
  Germany}.
\newblock
\newblock


\bibitem[Eapen and Adhithyan({[n.\,d.]})]%
        {eapenpersonalization}
\bibfield{author}{\bibinfo{person}{Joel Eapen} {and} \bibinfo{person}{VS
  Adhithyan}.} \bibinfo{year}{[n.\,d.]}\natexlab{}.
\newblock \showarticletitle{Personalization and Customization of LLM
  Responses}.
\newblock  (\bibinfo{year}{[n.\,d.]}).
\newblock


\bibitem[Ehsan et~al\mbox{.}(2021)]%
        {ehsan2021expanding}
\bibfield{author}{\bibinfo{person}{Upol Ehsan}, \bibinfo{person}{Q~Vera Liao},
  \bibinfo{person}{Michael Muller}, \bibinfo{person}{Mark~O Riedl}, {and}
  \bibinfo{person}{Justin~D Weisz}.} \bibinfo{year}{2021}\natexlab{}.
\newblock \showarticletitle{Expanding explainability: Towards social
  transparency in ai systems}. In \bibinfo{booktitle}{\emph{Proceedings of the
  2021 CHI Conference on Human Factors in Computing Systems}}.
  \bibinfo{pages}{1--19}.
\newblock


\bibitem[Fast et~al\mbox{.}(2018)]%
        {fast2018iris}
\bibfield{author}{\bibinfo{person}{Ethan Fast}, \bibinfo{person}{Binbin Chen},
  \bibinfo{person}{Julia Mendelsohn}, \bibinfo{person}{Jonathan Bassen}, {and}
  \bibinfo{person}{Michael~S Bernstein}.} \bibinfo{year}{2018}\natexlab{}.
\newblock \showarticletitle{Iris: A conversational agent for complex tasks}. In
  \bibinfo{booktitle}{\emph{Proceedings of the 2018 CHI conference on human
  factors in computing systems}}. \bibinfo{pages}{1--12}.
\newblock


\bibitem[Gauthier et~al\mbox{.}(2022)]%
        {gauthier2022will}
\bibfield{author}{\bibinfo{person}{Robert~P Gauthier},
  \bibinfo{person}{Mary~Jean Costello}, {and} \bibinfo{person}{James~R
  Wallace}.} \bibinfo{year}{2022}\natexlab{}.
\newblock \showarticletitle{“I Will Not Drink With You Today”: A
  Topic-Guided Thematic Analysis of Addiction Recovery on Reddit}. In
  \bibinfo{booktitle}{\emph{Proceedings of the 2022 CHI Conference on Human
  Factors in Computing Systems}}. \bibinfo{pages}{1--17}.
\newblock


\bibitem[Gehring et~al\mbox{.}(2012)]%
        {10.1145/2166966.2166985}
\bibfield{author}{\bibinfo{person}{Sven Gehring}, \bibinfo{person}{Markus
  L\"{o}chtefeld}, \bibinfo{person}{Florian Daiber}, \bibinfo{person}{Matthias
  B\"{o}hmer}, {and} \bibinfo{person}{Antonio Kr\"{u}ger}.}
  \bibinfo{year}{2012}\natexlab{}.
\newblock \showarticletitle{Using Intelligent Natural User Interfaces to
  Support Sales Conversations}. In \bibinfo{booktitle}{\emph{Proceedings of the
  2012 ACM International Conference on Intelligent User Interfaces}} (Lisbon,
  Portugal) \emph{(\bibinfo{series}{IUI '12})}. \bibinfo{publisher}{Association
  for Computing Machinery}, \bibinfo{address}{New York, NY, USA},
  \bibinfo{pages}{97–100}.
\newblock
\showISBNx{9781450310482}
\urldef\tempurl%
\url{https://doi.org/10.1145/2166966.2166985}
\showDOI{\tempurl}


\bibitem[Gkinko and Elbanna(2020)]%
        {gkinko2020creation}
\bibfield{author}{\bibinfo{person}{L Gkinko} {and} \bibinfo{person}{A
  Elbanna}.} \bibinfo{year}{2020}\natexlab{}.
\newblock \showarticletitle{The creation of chatbots at work: an organizational
  perspective}.
\newblock \bibinfo{journal}{\emph{AI@ Work. ai. reshapingwork. net, Amsterdam}}
  (\bibinfo{year}{2020}), \bibinfo{pages}{5--6}.
\newblock


\bibitem[Gkinko and Elbanna(2023)]%
        {gkinko2023appropriation}
\bibfield{author}{\bibinfo{person}{Lorentsa Gkinko} {and}
  \bibinfo{person}{Amany Elbanna}.} \bibinfo{year}{2023}\natexlab{}.
\newblock \showarticletitle{The appropriation of conversational AI in the
  workplace: A taxonomy of AI chatbot users}.
\newblock  (\bibinfo{year}{2023}).
\newblock


\bibitem[Gmeiner and Yildirim(2023)]%
        {gmeiner2023dimensions}
\bibfield{author}{\bibinfo{person}{Frederic Gmeiner} {and} \bibinfo{person}{Nur
  Yildirim}.} \bibinfo{year}{2023}\natexlab{}.
\newblock \showarticletitle{Dimensions for Designing LLM-based Writing
  Support}. In \bibinfo{booktitle}{\emph{In2Writing Workshop at CHI}}.
\newblock


\bibitem[Greene and Myerson(2011)]%
        {greene2011space}
\bibfield{author}{\bibinfo{person}{Catherine Greene} {and}
  \bibinfo{person}{Jeremy Myerson}.} \bibinfo{year}{2011}\natexlab{}.
\newblock \showarticletitle{Space for thought: designing for knowledge
  workers}.
\newblock \bibinfo{journal}{\emph{Facilities}} \bibinfo{volume}{29},
  \bibinfo{number}{1/2} (\bibinfo{year}{2011}), \bibinfo{pages}{19--30}.
\newblock


\bibitem[Grover et~al\mbox{.}(2020)]%
        {10.1145/3377325.3377507}
\bibfield{author}{\bibinfo{person}{Ted Grover}, \bibinfo{person}{Kael Rowan},
  \bibinfo{person}{Jina Suh}, \bibinfo{person}{Daniel McDuff}, {and}
  \bibinfo{person}{Mary Czerwinski}.} \bibinfo{year}{2020}\natexlab{}.
\newblock \showarticletitle{Design and Evaluation of Intelligent Agent
  Prototypes for Assistance with Focus and Productivity at Work}. In
  \bibinfo{booktitle}{\emph{Proceedings of the 25th International Conference on
  Intelligent User Interfaces}} (Cagliari, Italy) \emph{(\bibinfo{series}{IUI
  '20})}. \bibinfo{publisher}{Association for Computing Machinery},
  \bibinfo{address}{New York, NY, USA}, \bibinfo{pages}{390–400}.
\newblock
\showISBNx{9781450371186}
\urldef\tempurl%
\url{https://doi.org/10.1145/3377325.3377507}
\showDOI{\tempurl}


\bibitem[Guillou et~al\mbox{.}(2020)]%
        {guillou2020your}
\bibfield{author}{\bibinfo{person}{Hayley Guillou}, \bibinfo{person}{Kevin
  Chow}, \bibinfo{person}{Thomas Fritz}, {and} \bibinfo{person}{Joanna
  McGrenere}.} \bibinfo{year}{2020}\natexlab{}.
\newblock \showarticletitle{Is your time well spent? reflecting on knowledge
  work more holistically}. In \bibinfo{booktitle}{\emph{Proceedings of the 2020
  CHI Conference on Human Factors in Computing Systems}}.
  \bibinfo{pages}{1--9}.
\newblock


\bibitem[Haque et~al\mbox{.}(2022)]%
        {haque2022exploring}
\bibfield{author}{\bibinfo{person}{Mubin~Ul Haque}, \bibinfo{person}{Isuru
  Dharmadasa}, \bibinfo{person}{Zarrin~Tasnim Sworna},
  \bibinfo{person}{Roshan~Namal Rajapakse}, {and} \bibinfo{person}{Hussain
  Ahmad}.} \bibinfo{year}{2022}\natexlab{}.
\newblock \bibinfo{title}{"I think this is the most disruptive technology":
  Exploring Sentiments of ChatGPT Early Adopters using Twitter Data}.
\newblock
\newblock


\bibitem[He et~al\mbox{.}(2025)]%
        {he2025contributionsdeservecreditperceptions}
\bibfield{author}{\bibinfo{person}{Jessica He}, \bibinfo{person}{Stephanie
  Houde}, {and} \bibinfo{person}{Justin~D. Weisz}.}
  \bibinfo{year}{2025}\natexlab{}.
\newblock \bibinfo{title}{Which Contributions Deserve Credit? Perceptions of
  Attribution in Human-AI Co-Creation}.
\newblock
\newblock
\showeprint[arxiv]{2502.18357}~[cs.HC]
\urldef\tempurl%
\url{https://arxiv.org/abs/2502.18357}
\showURL{%
\tempurl}


\bibitem[He et~al\mbox{.}(2023)]%
        {he2023understanding}
\bibfield{author}{\bibinfo{person}{Jessica He}, \bibinfo{person}{David
  Piorkowski}, \bibinfo{person}{Michael Muller}, \bibinfo{person}{Kristina
  Brimijoin}, \bibinfo{person}{Stephanie Houde}, {and}
  \bibinfo{person}{Justin~D Weisz}.} \bibinfo{year}{2023}\natexlab{}.
\newblock \showarticletitle{Understanding How Task Dimensions Impact Automation
  Preferences with a Conversational Task Assistant}.
\newblock


\bibitem[Houben et~al\mbox{.}(2013)]%
        {houben2013activity}
\bibfield{author}{\bibinfo{person}{Steven Houben}, \bibinfo{person}{Jakob~E
  Bardram}, \bibinfo{person}{Jo Vermeulen}, \bibinfo{person}{Kris Luyten},
  {and} \bibinfo{person}{Karin Coninx}.} \bibinfo{year}{2013}\natexlab{}.
\newblock \showarticletitle{Activity-centric support for ad hoc knowledge work:
  A case study of co-activity manager}. In
  \bibinfo{booktitle}{\emph{Proceedings of the SIGCHI Conference on Human
  Factors in Computing Systems}}. \bibinfo{pages}{2263--2272}.
\newblock


\bibitem[Hu and Lee(2022)]%
        {hu2022scrapbook}
\bibfield{author}{\bibinfo{person}{Donghan Hu} {and} \bibinfo{person}{Sang~Won
  Lee}.} \bibinfo{year}{2022}\natexlab{}.
\newblock \showarticletitle{Scrapbook: Screenshot-Based Bookmarks for Effective
  Digital Resource Curation across Applications}. In
  \bibinfo{booktitle}{\emph{Proceedings of the 35th Annual ACM Symposium on
  User Interface Software and Technology}}. \bibinfo{pages}{1--13}.
\newblock


\bibitem[Iivari and Linger(1999)]%
        {iivari1999knowledge}
\bibfield{author}{\bibinfo{person}{Juhani Iivari} {and} \bibinfo{person}{Henry
  Linger}.} \bibinfo{year}{1999}\natexlab{}.
\newblock \showarticletitle{Knowledge work as collaborative work: A situated
  activity theory view}. In \bibinfo{booktitle}{\emph{Proceedings of the 32nd
  Annual Hawaii International Conference on Systems Sciences. 1999. HICSS-32.
  Abstracts and CD-ROM of Full Papers}}. IEEE, \bibinfo{pages}{10--pp}.
\newblock


\bibitem[Jessup et~al\mbox{.}(2019)]%
        {jessup2019measurement}
\bibfield{author}{\bibinfo{person}{Sarah~A Jessup}, \bibinfo{person}{Tamera~R
  Schneider}, \bibinfo{person}{Gene~M Alarcon}, \bibinfo{person}{Tyler~J Ryan},
  {and} \bibinfo{person}{August Capiola}.} \bibinfo{year}{2019}\natexlab{}.
\newblock \showarticletitle{The measurement of the propensity to trust
  automation}. In \bibinfo{booktitle}{\emph{Virtual, Augmented and Mixed
  Reality. Applications and Case Studies: 11th International Conference, VAMR
  2019, Held as Part of the 21st HCI International Conference, HCII 2019,
  Orlando, FL, USA, July 26--31, 2019, Proceedings, Part II 21}}. Springer,
  \bibinfo{pages}{476--489}.
\newblock


\bibitem[Kacperski et~al\mbox{.}(2023)]%
        {kacperski2023users}
\bibfield{author}{\bibinfo{person}{Celina Kacperski}, \bibinfo{person}{Roberto
  Ulloa}, \bibinfo{person}{Denis Bonnay}, \bibinfo{person}{Juhi Kulshrestha},
  \bibinfo{person}{Peter Selb}, {and} \bibinfo{person}{Andreas Spitz}.}
  \bibinfo{year}{2023}\natexlab{}.
\newblock \showarticletitle{Who are the users of ChatGPT? Implications for the
  digital divide from web tracking data}.
\newblock \bibinfo{journal}{\emph{arXiv preprint arXiv:2309.02142}}
  (\bibinfo{year}{2023}).
\newblock


\bibitem[Kaptelinin et~al\mbox{.}(1999)]%
        {kaptelinin1999methods}
\bibfield{author}{\bibinfo{person}{Victor Kaptelinin},
  \bibinfo{person}{Bonnie~A Nardi}, {and} \bibinfo{person}{Catriona Macaulay}.}
  \bibinfo{year}{1999}\natexlab{}.
\newblock \showarticletitle{Methods \& tools: The activity checklist: a tool
  for representing the “space” of context}.
\newblock \bibinfo{journal}{\emph{interactions}} \bibinfo{volume}{6},
  \bibinfo{number}{4} (\bibinfo{year}{1999}), \bibinfo{pages}{27--39}.
\newblock


\bibitem[Khosrawi-Rad et~al\mbox{.}(2022)]%
        {khosrawi2022conversational}
\bibfield{author}{\bibinfo{person}{Bijan Khosrawi-Rad}, \bibinfo{person}{Heidi
  Rinn}, \bibinfo{person}{Ricarda Schlimbach}, \bibinfo{person}{Pia Gebbing},
  \bibinfo{person}{Xingyue Yang}, \bibinfo{person}{Christoph Lattemann},
  \bibinfo{person}{Daniel Markgraf}, {and} \bibinfo{person}{Susanne
  Robra-Bissantz}.} \bibinfo{year}{2022}\natexlab{}.
\newblock \showarticletitle{Conversational agents in education--a systematic
  literature review}.
\newblock  (\bibinfo{year}{2022}).
\newblock


\bibitem[Ko et~al\mbox{.}(2023)]%
        {10.1145/3581641.3584078}
\bibfield{author}{\bibinfo{person}{Hyung-Kwon Ko}, \bibinfo{person}{Gwanmo
  Park}, \bibinfo{person}{Hyeon Jeon}, \bibinfo{person}{Jaemin Jo},
  \bibinfo{person}{Juho Kim}, {and} \bibinfo{person}{Jinwook Seo}.}
  \bibinfo{year}{2023}\natexlab{}.
\newblock \showarticletitle{Large-Scale Text-to-Image Generation Models for
  Visual Artists’ Creative Works}. In \bibinfo{booktitle}{\emph{Proceedings
  of the 28th International Conference on Intelligent User Interfaces}}
  (Sydney, NSW, Australia) \emph{(\bibinfo{series}{IUI '23})}.
  \bibinfo{publisher}{Association for Computing Machinery},
  \bibinfo{address}{New York, NY, USA}, \bibinfo{pages}{919–933}.
\newblock
\showISBNx{9798400701061}
\urldef\tempurl%
\url{https://doi.org/10.1145/3581641.3584078}
\showDOI{\tempurl}


\bibitem[Kogan and Muller(2006)]%
        {kogan2006ethnographic}
\bibfield{author}{\bibinfo{person}{Sandra~L Kogan} {and}
  \bibinfo{person}{Michael~J Muller}.} \bibinfo{year}{2006}\natexlab{}.
\newblock \showarticletitle{Ethnographic study of collaborative knowledge
  work}.
\newblock \bibinfo{journal}{\emph{IBM Systems Journal}} \bibinfo{volume}{45},
  \bibinfo{number}{4} (\bibinfo{year}{2006}), \bibinfo{pages}{759--771}.
\newblock


\bibitem[Lai et~al\mbox{.}(2014)]%
        {10.1145/2557500.2557539}
\bibfield{author}{\bibinfo{person}{Jennifer Lai}, \bibinfo{person}{Jie Lu},
  \bibinfo{person}{Shimei Pan}, \bibinfo{person}{Danny Soroker},
  \bibinfo{person}{Mercan Topkara}, \bibinfo{person}{Justin Weisz},
  \bibinfo{person}{Jeff Boston}, {and} \bibinfo{person}{Jason Crawford}.}
  \bibinfo{year}{2014}\natexlab{}.
\newblock \showarticletitle{Expediting Expertise: Supporting Informal Social
  Learning in the Enterprise}. In \bibinfo{booktitle}{\emph{Proceedings of the
  19th International Conference on Intelligent User Interfaces}} (Haifa,
  Israel) \emph{(\bibinfo{series}{IUI '14})}. \bibinfo{publisher}{Association
  for Computing Machinery}, \bibinfo{address}{New York, NY, USA},
  \bibinfo{pages}{133–142}.
\newblock
\showISBNx{9781450321846}
\urldef\tempurl%
\url{https://doi.org/10.1145/2557500.2557539}
\showDOI{\tempurl}


\bibitem[Laranjo et~al\mbox{.}(2018)]%
        {laranjo2018conversational}
\bibfield{author}{\bibinfo{person}{Liliana Laranjo}, \bibinfo{person}{Adam~G
  Dunn}, \bibinfo{person}{Huong~Ly Tong}, \bibinfo{person}{Ahmet~Baki
  Kocaballi}, \bibinfo{person}{Jessica Chen}, \bibinfo{person}{Rabia Bashir},
  \bibinfo{person}{Didi Surian}, \bibinfo{person}{Blanca Gallego},
  \bibinfo{person}{Farah Magrabi}, \bibinfo{person}{Annie~YS Lau},
  {et~al\mbox{.}}} \bibinfo{year}{2018}\natexlab{}.
\newblock \showarticletitle{Conversational agents in healthcare: a systematic
  review}.
\newblock \bibinfo{journal}{\emph{Journal of the American Medical Informatics
  Association}} \bibinfo{volume}{25}, \bibinfo{number}{9}
  (\bibinfo{year}{2018}), \bibinfo{pages}{1248--1258}.
\newblock


\bibitem[Li et~al\mbox{.}(2023b)]%
        {li2023teach}
\bibfield{author}{\bibinfo{person}{Cheng Li}, \bibinfo{person}{Mingyang Zhang},
  \bibinfo{person}{Qiaozhu Mei}, \bibinfo{person}{Yaqing Wang},
  \bibinfo{person}{Spurthi~Amba Hombaiah}, \bibinfo{person}{Yi Liang}, {and}
  \bibinfo{person}{Michael Bendersky}.} \bibinfo{year}{2023}\natexlab{b}.
\newblock \showarticletitle{Teach LLMs to Personalize--An Approach inspired by
  Writing Education}.
\newblock \bibinfo{journal}{\emph{arXiv preprint arXiv:2308.07968}}
  (\bibinfo{year}{2023}).
\newblock


\bibitem[Li et~al\mbox{.}(2023a)]%
        {li2023chatgpt}
\bibfield{author}{\bibinfo{person}{Lingyao Li}, \bibinfo{person}{Zihui Ma},
  \bibinfo{person}{Lizhou Fan}, \bibinfo{person}{Sanggyu Lee},
  \bibinfo{person}{Huizi Yu}, {and} \bibinfo{person}{Libby Hemphill}.}
  \bibinfo{year}{2023}\natexlab{a}.
\newblock \bibinfo{title}{ChatGPT in education: A discourse analysis of worries
  and concerns on social media}.
\newblock
\newblock


\bibitem[Liao et~al\mbox{.}(2023)]%
        {liao2023designerly}
\bibfield{author}{\bibinfo{person}{Q~Vera Liao}, \bibinfo{person}{Hariharan
  Subramonyam}, \bibinfo{person}{Jennifer Wang}, {and}
  \bibinfo{person}{Jennifer Wortman~Vaughan}.} \bibinfo{year}{2023}\natexlab{}.
\newblock \showarticletitle{Designerly understanding: Information needs for
  model transparency to support design ideation for AI-powered user
  experience}. In \bibinfo{booktitle}{\emph{Proceedings of the 2023 CHI
  conference on human factors in computing systems}}. \bibinfo{pages}{1--21}.
\newblock


\bibitem[Liao and Vaughan(2023)]%
        {liao2023ai}
\bibfield{author}{\bibinfo{person}{Q~Vera Liao} {and}
  \bibinfo{person}{Jennifer~Wortman Vaughan}.} \bibinfo{year}{2023}\natexlab{}.
\newblock \showarticletitle{AI Transparency in the Age of LLMs: A
  Human-Centered Research Roadmap}.
\newblock \bibinfo{journal}{\emph{arXiv preprint arXiv:2306.01941}}
  (\bibinfo{year}{2023}).
\newblock


\bibitem[Lu et~al\mbox{.}(2011)]%
        {10.1145/1943403.1943434}
\bibfield{author}{\bibinfo{person}{Jie Lu}, \bibinfo{person}{Shimei Pan},
  \bibinfo{person}{Jennifer~C. Lai}, {and} \bibinfo{person}{Zhen Wen}.}
  \bibinfo{year}{2011}\natexlab{}.
\newblock \showarticletitle{Information at Your Fingertips: Contextual IR in
  Enterprise Email}. In \bibinfo{booktitle}{\emph{Proceedings of the 16th
  International Conference on Intelligent User Interfaces}} (Palo Alto, CA,
  USA) \emph{(\bibinfo{series}{IUI '11})}. \bibinfo{publisher}{Association for
  Computing Machinery}, \bibinfo{address}{New York, NY, USA},
  \bibinfo{pages}{205–214}.
\newblock
\showISBNx{9781450304191}
\urldef\tempurl%
\url{https://doi.org/10.1145/1943403.1943434}
\showDOI{\tempurl}


\bibitem[Makkonen et~al\mbox{.}(2023)]%
        {makkonen2023effects}
\bibfield{author}{\bibinfo{person}{Markus Makkonen}, \bibinfo{person}{Markus
  Salo}, {and} \bibinfo{person}{Henri Pirkkalainen}.}
  \bibinfo{year}{2023}\natexlab{}.
\newblock \showarticletitle{The Effects of Job and User Characteristics on the
  Perceived Usefulness and Use Continuance Intention of Generative Artificial
  Intelligence Chatbots at Work}.
\newblock \bibinfo{journal}{\emph{Proceedings http://ceur-ws. org ISSN}}
  \bibinfo{volume}{1613} (\bibinfo{year}{2023}), \bibinfo{pages}{0073}.
\newblock


\bibitem[Mark et~al\mbox{.}(2005)]%
        {mark2005no}
\bibfield{author}{\bibinfo{person}{Gloria Mark}, \bibinfo{person}{Victor~M
  Gonzalez}, {and} \bibinfo{person}{Justin Harris}.}
  \bibinfo{year}{2005}\natexlab{}.
\newblock \showarticletitle{No task left behind? Examining the nature of
  fragmented work}. In \bibinfo{booktitle}{\emph{Proceedings of the SIGCHI
  conference on Human factors in computing systems}}.
  \bibinfo{pages}{321--330}.
\newblock


\bibitem[Mark et~al\mbox{.}(2008)]%
        {mark2008cost}
\bibfield{author}{\bibinfo{person}{Gloria Mark}, \bibinfo{person}{Daniela
  Gudith}, {and} \bibinfo{person}{Ulrich Klocke}.}
  \bibinfo{year}{2008}\natexlab{}.
\newblock \showarticletitle{The cost of interrupted work: more speed and
  stress}. In \bibinfo{booktitle}{\emph{Proceedings of the SIGCHI conference on
  Human Factors in Computing Systems}}. \bibinfo{pages}{107--110}.
\newblock


\bibitem[Miller and Cooper(2022)]%
        {miller2022barriers}
\bibfield{author}{\bibinfo{person}{Josh~Aaron Miller} {and}
  \bibinfo{person}{Seth Cooper}.} \bibinfo{year}{2022}\natexlab{}.
\newblock \showarticletitle{Barriers to expertise in citizen science games}. In
  \bibinfo{booktitle}{\emph{Proceedings of the 2022 CHI Conference on Human
  Factors in Computing Systems}}. \bibinfo{pages}{1--25}.
\newblock


\bibitem[Mitchell et~al\mbox{.}(2019)]%
        {mitchell2019model}
\bibfield{author}{\bibinfo{person}{Margaret Mitchell}, \bibinfo{person}{Simone
  Wu}, \bibinfo{person}{Andrew Zaldivar}, \bibinfo{person}{Parker Barnes},
  \bibinfo{person}{Lucy Vasserman}, \bibinfo{person}{Ben Hutchinson},
  \bibinfo{person}{Elena Spitzer}, \bibinfo{person}{Inioluwa~Deborah Raji},
  {and} \bibinfo{person}{Timnit Gebru}.} \bibinfo{year}{2019}\natexlab{}.
\newblock \showarticletitle{Model cards for model reporting}. In
  \bibinfo{booktitle}{\emph{Proceedings of the conference on fairness,
  accountability, and transparency}}. \bibinfo{pages}{220--229}.
\newblock


\bibitem[Muller et~al\mbox{.}(2024)]%
        {muller2024group}
\bibfield{author}{\bibinfo{person}{Michael Muller}, \bibinfo{person}{Stephanie
  Houde}, \bibinfo{person}{Gabriel Gonzalez}, \bibinfo{person}{Kristina
  Brimijoin}, \bibinfo{person}{Steven~I Ross}, \bibinfo{person}{Dario
  Andres~Silva Moran}, {and} \bibinfo{person}{Justin~D Weisz}.}
  \bibinfo{year}{2024}\natexlab{}.
\newblock \showarticletitle{Group Brainstorming with an AI Agent: Creating and
  Selecting Ideas}. In \bibinfo{booktitle}{\emph{International Conference on
  Computational Creativity}}.
\newblock


\bibitem[Mundbrod et~al\mbox{.}(2013)]%
        {mundbrod2013towards}
\bibfield{author}{\bibinfo{person}{Nicolas Mundbrod}, \bibinfo{person}{Jens
  Kolb}, {and} \bibinfo{person}{Manfred Reichert}.}
  \bibinfo{year}{2013}\natexlab{}.
\newblock \showarticletitle{Towards a system support of collaborative knowledge
  work}. In \bibinfo{booktitle}{\emph{Business Process Management Workshops:
  BPM 2012 International Workshops, Tallinn, Estonia, September 3, 2012.
  Revised Papers 10}}. Springer, \bibinfo{pages}{31--42}.
\newblock


\bibitem[Palvalin(2019)]%
        {palvalin2019matters}
\bibfield{author}{\bibinfo{person}{Miikka Palvalin}.}
  \bibinfo{year}{2019}\natexlab{}.
\newblock \showarticletitle{What matters for knowledge work productivity?}
\newblock \bibinfo{journal}{\emph{Employee Relations}} \bibinfo{volume}{41},
  \bibinfo{number}{1} (\bibinfo{year}{2019}), \bibinfo{pages}{209--227}.
\newblock


\bibitem[Park et~al\mbox{.}(2021)]%
        {park2021human}
\bibfield{author}{\bibinfo{person}{Hyanghee Park}, \bibinfo{person}{Daehwan
  Ahn}, \bibinfo{person}{Kartik Hosanagar}, {and} \bibinfo{person}{Joonhwan
  Lee}.} \bibinfo{year}{2021}\natexlab{}.
\newblock \showarticletitle{Human-AI interaction in human resource management:
  Understanding why employees resist algorithmic evaluation at workplaces and
  how to mitigate burdens}. In \bibinfo{booktitle}{\emph{Proceedings of the
  2021 CHI Conference on Human Factors in Computing Systems}}.
  \bibinfo{pages}{1--15}.
\newblock


\bibitem[Parker and Grote(2022)]%
        {parker2022automation}
\bibfield{author}{\bibinfo{person}{Sharon~K Parker} {and}
  \bibinfo{person}{Gudela Grote}.} \bibinfo{year}{2022}\natexlab{}.
\newblock \showarticletitle{Automation, algorithms, and beyond: Why work design
  matters more than ever in a digital world}.
\newblock \bibinfo{journal}{\emph{Applied Psychology}} \bibinfo{volume}{71},
  \bibinfo{number}{4} (\bibinfo{year}{2022}), \bibinfo{pages}{1171--1204}.
\newblock


\bibitem[Praveen and Vajrobol(2023)]%
        {praveen2023understanding}
\bibfield{author}{\bibinfo{person}{S.~V. Praveen} {and}
  \bibinfo{person}{Vajratiya Vajrobol}.} \bibinfo{year}{2023}\natexlab{}.
\newblock \bibinfo{title}{Understanding the Perceptions of Healthcare
  Researchers Regarding ChatGPT: A Study Based on Bidirectional Encoder
  Representation from Transformers (BERT) Sentiment Analysis and Topic
  Modeling}.
\newblock
\newblock


\bibitem[Renney et~al\mbox{.}(2022)]%
        {renney2022studying}
\bibfield{author}{\bibinfo{person}{Nathan Renney}, \bibinfo{person}{Benedict
  Gaster}, \bibinfo{person}{Tom Mitchell}, {and} \bibinfo{person}{Harri
  Renney}.} \bibinfo{year}{2022}\natexlab{}.
\newblock \showarticletitle{Studying How Digital Luthiers Choose Their Tools}.
  In \bibinfo{booktitle}{\emph{Proceedings of the 2022 CHI Conference on Human
  Factors in Computing Systems}}. \bibinfo{pages}{1--18}.
\newblock


\bibitem[Ritala et~al\mbox{.}(2023)]%
        {ritala2023transforming}
\bibfield{author}{\bibinfo{person}{Paavo Ritala}, \bibinfo{person}{Mika
  Ruokonen}, {and} \bibinfo{person}{Laavanya Ramaul}.}
  \bibinfo{year}{2023}\natexlab{}.
\newblock \showarticletitle{Transforming boundaries: how does ChatGPT change
  knowledge work?}
\newblock \bibinfo{journal}{\emph{Journal of Business Strategy}}
  (\bibinfo{year}{2023}).
\newblock


\bibitem[Ross et~al\mbox{.}(2023)]%
        {ross2023programmer}
\bibfield{author}{\bibinfo{person}{Steven~I Ross}, \bibinfo{person}{Fernando
  Martinez}, \bibinfo{person}{Stephanie Houde}, \bibinfo{person}{Michael
  Muller}, {and} \bibinfo{person}{Justin~D Weisz}.}
  \bibinfo{year}{2023}\natexlab{}.
\newblock \showarticletitle{The programmer’s assistant: Conversational
  interaction with a large language model for software development}. In
  \bibinfo{booktitle}{\emph{Proceedings of the 28th International Conference on
  Intelligent User Interfaces}}. \bibinfo{pages}{491--514}.
\newblock


\bibitem[Shaer et~al\mbox{.}(2024)]%
        {shaer2024toward}
\bibfield{author}{\bibinfo{person}{Orit Shaer}, \bibinfo{person}{Angelora
  Cooper}, \bibinfo{person}{Andrew~L Kun}, {and} \bibinfo{person}{Osnat
  Mokryn}.} \bibinfo{year}{2024}\natexlab{}.
\newblock \showarticletitle{Toward Enhancing Ideation through Collaborative
  Group-AI Brainwriting.}. In \bibinfo{booktitle}{\emph{IUI Workshops}}.
\newblock


\bibitem[Shen et~al\mbox{.}(2008)]%
        {shen2008automatically}
\bibfield{author}{\bibinfo{person}{Jianqiang Shen}, \bibinfo{person}{Werner
  Geyer}, \bibinfo{person}{Michael Muller}, \bibinfo{person}{Casey Dugan},
  \bibinfo{person}{Beth Brownholtz}, {and} \bibinfo{person}{David~R Millen}.}
  \bibinfo{year}{2008}\natexlab{}.
\newblock \showarticletitle{Automatically finding and recommending resources to
  support knowledge workers' activities}. In
  \bibinfo{booktitle}{\emph{Proceedings of the 13th international conference on
  Intelligent user interfaces}}. \bibinfo{pages}{207--216}.
\newblock


\bibitem[Shen et~al\mbox{.}(2023)]%
        {shen2023chatgpt}
\bibfield{author}{\bibinfo{person}{Xinyue Shen}, \bibinfo{person}{Zeyuan Chen},
  \bibinfo{person}{Michael Backes}, {and} \bibinfo{person}{Yang Zhang}.}
  \bibinfo{year}{2023}\natexlab{}.
\newblock \bibinfo{title}{In ChatGPT We Trust? Measuring and Characterizing the
  Reliability of ChatGPT}.
\newblock
\newblock


\bibitem[Shoufan(2023)]%
        {shoufan2023exploring}
\bibfield{author}{\bibinfo{person}{Abdulhadi Shoufan}.}
  \bibinfo{year}{2023}\natexlab{}.
\newblock \bibinfo{title}{Exploring Students’ Perceptions of ChatGPT:
  Thematic Analysis and Follow-Up Survey}.
\newblock
\newblock


\bibitem[Siegmund et~al\mbox{.}(2014)]%
        {siegmund2014measuring}
\bibfield{author}{\bibinfo{person}{Janet Siegmund}, \bibinfo{person}{Christian
  K{\"a}stner}, \bibinfo{person}{J{\"o}rg Liebig}, \bibinfo{person}{Sven Apel},
  {and} \bibinfo{person}{Stefan Hanenberg}.} \bibinfo{year}{2014}\natexlab{}.
\newblock \showarticletitle{Measuring and modeling programming experience}.
\newblock \bibinfo{journal}{\emph{Empirical Software Engineering}}
  \bibinfo{volume}{19} (\bibinfo{year}{2014}), \bibinfo{pages}{1299--1334}.
\newblock


\bibitem[Skjuve(2023)]%
        {skjuve4376834people}
\bibfield{author}{\bibinfo{person}{Marita Skjuve}.}
  \bibinfo{year}{2023}\natexlab{}.
\newblock \showarticletitle{Why people use chatgpt}.
\newblock \bibinfo{journal}{\emph{Available at SSRN 4376834}}
  (\bibinfo{year}{2023}).
\newblock


\bibitem[Skjuve et~al\mbox{.}(2023)]%
        {skjuve2023user}
\bibfield{author}{\bibinfo{person}{Marita Skjuve}, \bibinfo{person}{Asbj{\o}rn
  F{\o}lstad}, {and} \bibinfo{person}{Petter~Bae Brandtzaeg}.}
  \bibinfo{year}{2023}\natexlab{}.
\newblock \showarticletitle{The User Experience of ChatGPT: Findings from a
  Questionnaire Study of Early Users}. In \bibinfo{booktitle}{\emph{Proceedings
  of the 5th International Conference on Conversational User Interfaces}}.
  \bibinfo{pages}{1--10}.
\newblock


\bibitem[Tabor et~al\mbox{.}(2021)]%
        {10.1145/3411764.3445388}
\bibfield{author}{\bibinfo{person}{Aaron Tabor}, \bibinfo{person}{Scott
  Bateman}, \bibinfo{person}{Erik Scheme}, \bibinfo{person}{Book Sadprasid},
  {and} \bibinfo{person}{m.c. schraefel}.} \bibinfo{year}{2021}\natexlab{}.
\newblock \showarticletitle{Understanding the Design and Effectiveness of
  Peripheral Breathing Guide Use During Information Work}. In
  \bibinfo{booktitle}{\emph{Proceedings of the 2021 CHI Conference on Human
  Factors in Computing Systems}} (Yokohama, Japan) \emph{(\bibinfo{series}{CHI
  '21})}. \bibinfo{publisher}{Association for Computing Machinery},
  \bibinfo{address}{New York, NY, USA}, Article \bibinfo{articleno}{516},
  \bibinfo{numpages}{13}~pages.
\newblock
\showISBNx{9781450380966}
\urldef\tempurl%
\url{https://doi.org/10.1145/3411764.3445388}
\showDOI{\tempurl}


\bibitem[Taecharungroj(2023)]%
        {taecharungroj2023can}
\bibfield{author}{\bibinfo{person}{Viriya Taecharungroj}.}
  \bibinfo{year}{2023}\natexlab{}.
\newblock \showarticletitle{“What Can ChatGPT Do?” Analyzing Early
  Reactions to the Innovative AI Chatbot on Twitter}.
\newblock \bibinfo{journal}{\emph{Big Data and Cognitive Computing}}
  \bibinfo{volume}{7}, \bibinfo{number}{1} (\bibinfo{year}{2023}),
  \bibinfo{pages}{35}.
\newblock


\bibitem[Van~der Aalst et~al\mbox{.}(2003)]%
        {van2003workflow}
\bibfield{author}{\bibinfo{person}{Wil~MP Van~der Aalst},
  \bibinfo{person}{Boudewijn~F Van~Dongen}, \bibinfo{person}{Joachim Herbst},
  \bibinfo{person}{Laura Maruster}, \bibinfo{person}{Guido Schimm}, {and}
  \bibinfo{person}{Anton~JMM Weijters}.} \bibinfo{year}{2003}\natexlab{}.
\newblock \showarticletitle{Workflow mining: A survey of issues and
  approaches}.
\newblock \bibinfo{journal}{\emph{Data \& knowledge engineering}}
  \bibinfo{volume}{47}, \bibinfo{number}{2} (\bibinfo{year}{2003}),
  \bibinfo{pages}{237--267}.
\newblock


\bibitem[Van~Elst et~al\mbox{.}(2003)]%
        {van2003weakly}
\bibfield{author}{\bibinfo{person}{Ludger Van~Elst}, \bibinfo{person}{F-R
  Aschoff}, \bibinfo{person}{Ansgar Bernardi}, {and} \bibinfo{person}{S
  Schwarz}.} \bibinfo{year}{2003}\natexlab{}.
\newblock \showarticletitle{Weakly-structured workflows for knowledge-intensive
  tasks: An experimental evaluation}. In \bibinfo{booktitle}{\emph{WET ICE
  2003. Proceedings. Twelfth IEEE International Workshops on Enabling
  Technologies: Infrastructure for Collaborative Enterprises, 2003.}} IEEE,
  \bibinfo{pages}{340--345}.
\newblock


\bibitem[Vtyurina et~al\mbox{.}(2017)]%
        {vtyurina2017exploring}
\bibfield{author}{\bibinfo{person}{Alexandra Vtyurina}, \bibinfo{person}{Denis
  Savenkov}, \bibinfo{person}{Eugene Agichtein}, {and}
  \bibinfo{person}{Charles~LA Clarke}.} \bibinfo{year}{2017}\natexlab{}.
\newblock \showarticletitle{Exploring conversational search with humans,
  assistants, and wizards}. In \bibinfo{booktitle}{\emph{Proceedings of the
  2017 chi conference extended abstracts on human factors in computing
  systems}}. \bibinfo{pages}{2187--2193}.
\newblock


\bibitem[Whittaker and Sidner(1996)]%
        {whittaker1996email}
\bibfield{author}{\bibinfo{person}{Steve Whittaker} {and}
  \bibinfo{person}{Candace Sidner}.} \bibinfo{year}{1996}\natexlab{}.
\newblock \showarticletitle{Email overload: exploring personal information
  management of email}. In \bibinfo{booktitle}{\emph{Proceedings of the SIGCHI
  conference on Human factors in computing systems}}.
  \bibinfo{pages}{276--283}.
\newblock


\bibitem[Wu et~al\mbox{.}(2022)]%
        {wu2022promptchainer}
\bibfield{author}{\bibinfo{person}{Tongshuang Wu}, \bibinfo{person}{Ellen
  Jiang}, \bibinfo{person}{Aaron Donsbach}, \bibinfo{person}{Jeff Gray},
  \bibinfo{person}{Alejandra Molina}, \bibinfo{person}{Michael Terry}, {and}
  \bibinfo{person}{Carrie~J Cai}.} \bibinfo{year}{2022}\natexlab{}.
\newblock \showarticletitle{Promptchainer: Chaining large language model
  prompts through visual programming}. In \bibinfo{booktitle}{\emph{CHI
  Conference on Human Factors in Computing Systems Extended Abstracts}}.
  \bibinfo{pages}{1--10}.
\newblock


\bibitem[Xu et~al\mbox{.}(2017)]%
        {xu2017new}
\bibfield{author}{\bibinfo{person}{Anbang Xu}, \bibinfo{person}{Zhe Liu},
  \bibinfo{person}{Yufan Guo}, \bibinfo{person}{Vibha Sinha}, {and}
  \bibinfo{person}{Rama Akkiraju}.} \bibinfo{year}{2017}\natexlab{}.
\newblock \showarticletitle{A new chatbot for customer service on social
  media}. In \bibinfo{booktitle}{\emph{Proceedings of the 2017 CHI conference
  on human factors in computing systems}}. \bibinfo{pages}{3506--3510}.
\newblock


\bibitem[Yan et~al\mbox{.}(2023)]%
        {yan2023practical}
\bibfield{author}{\bibinfo{person}{Lixiang Yan}, \bibinfo{person}{Lele Sha},
  \bibinfo{person}{Linxuan Zhao}, \bibinfo{person}{Yuheng Li},
  \bibinfo{person}{Roberto Martinez-Maldonado}, \bibinfo{person}{Guanliang
  Chen}, \bibinfo{person}{Xinyu Li}, \bibinfo{person}{Yueqiao Jin}, {and}
  \bibinfo{person}{Dragan Ga{\v{s}}evi{\'c}}.} \bibinfo{year}{2023}\natexlab{}.
\newblock \showarticletitle{Practical and ethical challenges of large language
  models in education: A systematic scoping review}.
\newblock \bibinfo{journal}{\emph{British Journal of Educational Technology}}
  (\bibinfo{year}{2023}).
\newblock


\bibitem[Yao et~al\mbox{.}(2023)]%
        {yao2023reactsynergizingreasoningacting}
\bibfield{author}{\bibinfo{person}{Shunyu Yao}, \bibinfo{person}{Jeffrey Zhao},
  \bibinfo{person}{Dian Yu}, \bibinfo{person}{Nan Du}, \bibinfo{person}{Izhak
  Shafran}, \bibinfo{person}{Karthik Narasimhan}, {and} \bibinfo{person}{Yuan
  Cao}.} \bibinfo{year}{2023}\natexlab{}.
\newblock \bibinfo{title}{ReAct: Synergizing Reasoning and Acting in Language
  Models}.
\newblock
\newblock
\showeprint[arxiv]{2210.03629}~[cs.CL]
\urldef\tempurl%
\url{https://arxiv.org/abs/2210.03629}
\showURL{%
\tempurl}


\bibitem[Zheng et~al\mbox{.}(2023)]%
        {zheng2023innovating}
\bibfield{author}{\bibinfo{person}{Yue Zheng}, \bibinfo{person}{Laduona Wang},
  \bibinfo{person}{Baijie Feng}, \bibinfo{person}{Ailin Zhao}, {and}
  \bibinfo{person}{Yijun Wu}.} \bibinfo{year}{2023}\natexlab{}.
\newblock \showarticletitle{Innovating Healthcare: the role of ChatGPT in
  streamlining hospital workflow in the future}.
\newblock \bibinfo{journal}{\emph{Annals of Biomedical Engineering}}
  (\bibinfo{year}{2023}), \bibinfo{pages}{1--4}.
\newblock


\end{thebibliography}

\section*{Appendix}

\begin{table}[h!]
    \caption{Participant demographics of all  Survey1 \color{black}participants (n=216). We list all job roles in the table with more than one participant listing them. Participants also listed several other roles, like content creation,  education, audit, communications, operations, strategy, content writing, contract preparation, procurement, and proposal management.}
    \begin{tabular}{c | c} 
    \hline
            \begin{tabular}[t]{lll}
                Work Location (select one)& \# & \%\\ \hline
                North America &136&63\%\\ 
                Europe &47&22\%\\ 
                Asia &22&10\%\\ 
                South America &5&2\%\\ 
                Oceania &3&1\%\\ 
                Africa &3&1\%\\ 
      
                \hline
                AI experience (select all that apply)& \# & \%\\ \hline
                Tried consumer AI tools  &151&70\%\\ 
                Use consumer AI systems regularly &92&43\%\\ 
                Closely follow AI news &86&40\%\\ 
                Some work experience/education &69&32\%\\ 
                Work related to AI &53&25\%\\ 
                Significant AI work experience &26&12\%\\ 
                Don't know much about it &17&8\%\\ 
    
            \end{tabular}
        &
            \begin{tabular}[t]{lll}
                Job Responsibility (select all that apply)& \# & \%\\ \hline
                Technical &97&45\%\\ 
                Design &41&19\%\\ 
                Sales &35&16\%\\ 
                Research & 28 & 13\% \\ 
                Analyst &27&12\%\\ 
                User Research &24&11\%\\ 
                Marketing &24&11\%\\ 
                Management &18&8\%\\
                Customer Service &15&7\%\\ 
                Product/project Management &12&6\%\\ 
                Administrative & 10 & 5\% \\ 
                Human Resources  &6&3\%\\ 
                Executive &4&2\%\\ 
                Finance & 3 & 1\% \\
                Copy editing & 3 & 1\% \\ 
                Consulting & 3 & 1\% \\ 
                Legal &2&1\%\\
                Accounting & 2 & 1\% \\ 
            \end{tabular}
            \\ \hline
    \end{tabular}
    \label{tab-demo}
\end{table}

\begin{table}[h!]

    \caption{Participant demographics of all participants for Survey2 (n=107). We list all job roles in the table with more than one participant listing them. Participants also listed several other roles, like Hardware Development \& Support, Legal, Site Reliability Engineer, Technical Services, Supply Chain, and Research. Note that due to policies, the work location and job responsibility questions required slight modifications from Survey1.}
    \begin{tabular}{c | c} 
    \hline
            \begin{tabular}[t]{lll}
                Work Location (select one)& \# & \%\\ \hline
                Americas &56&52\%\\ 
                EMEA (Europe, the Middle East, and Africa) &42&39\%\\ 
                APAC (Asia Pacific)   &9&8\%\\ 
      
                \hline
                AI experience (select all that apply)& \# & \%\\ \hline
                Tried consumer AI tools  &71&66\%\\ 
                Use consumer AI systems regularly &44&41\%\\ 
                Closely follow AI news &34&32\%\\ 
                Some work experience/education &47&44\%\\ 
                Work related to AI &31&29\%\\ 
                Significant AI work experience &8&4\%\\ 
                Don't know much about it &9&8\%\\ 
    
            \end{tabular}
        &
            \begin{tabular}[t]{lll}
                Job Responsibility (select all that apply)& \# & \%\\ \hline
                Software Development \& Support&   22& 21\%\\
                Consultant&                        13& 12\%\\
                Technical Sepcialist &             10& 9\%\\
                Finance   &                         8& 7\%\\
                Marketing \& Communications &       7& 7\%\\
                Enterprise Operations &             5&  5\%\\
                Sales           &                   5&  5\%\\
                Project Management  &               5&   5\%\\
                Architect        &                  5&  5\%\\
                Human Resources  &                   3& 3\%\\
                Design         &                     3& 3\% \\
                Information Technology \& Services&  3&  3\% \\
                Communications          &            3&  3\% \\
                Offering Management    &             3&  3\% \\
                Data Science          &              2&   3\% \\
                General Management    &              2&  3\% \\
            \end{tabular}
            \\ \hline
    \end{tabular}
    \label{tab:demo2}
\end{table}

\end{document}